\documentclass[aps,prd,10pt,twocolumn,superscriptaddress,floatfix,nofootinbib,amsmath,amssymb,altaffilletter,preprintnumbers]{revtex4-1}
\usepackage[colorlinks=true,pdfstartview=FitV,breaklinks=true]{hyperref}

\usepackage[utf8]{inputenc}
\usepackage{graphicx}
\usepackage{booktabs}
\usepackage{tabularx}
\usepackage{textcomp}
\usepackage{xspace}
\usepackage{amsmath}
\usepackage{amsfonts}
\usepackage{amssymb}
\usepackage{upgreek}
\usepackage[dvipsnames,table]{xcolor}
\usepackage[normalem]{ulem}
\usepackage{csquotes}

\hypersetup{urlcolor=BlueViolet,
	    citecolor=Plum,
	    linkcolor=PineGreen}
\usepackage[official]{eurosym}
\definecolor{lightgray}{rgb}{0.9,0.9,0.9}	    
\definecolor{green}{rgb}{0,0.5,0}
\definecolor{red}{rgb}{1,0,0}
\definecolor{blue}{rgb}{0,0,0.5}

\newcommand\codo[1]{{\tt #1}}
\newcommand{\n}{{\ensuremath{\mathfrak{n}}}\xspace}

\begin{document}

\title{Axion minivoids and implications for direct detection}

\author{Benedikt Eggemeier}
\affiliation{Institut f\"ur Astrophysik, Georg-August-Universit\"at G\"ottingen, D-37077 G\"ottingen, Germany}

\author{Ciaran A. J. O'Hare}
\affiliation{School of Physics, Physics Road, The University of Sydney, NSW 2006 Camperdown, Sydney, Australia}
\affiliation{ARC Centre of Excellence for Dark Matter Particle Physics, Sydney, Australia}

\author{Giovanni Pierobon}\email{g.pierobon@unsw.edu.au}
\affiliation{School of Physics, The University of New South Wales, NSW 2052 Kensington, Sydney, Australia}

\author{Javier Redondo}
\affiliation{CAPA \& Departamento de Fisica Teorica, Universidad de Zaragoza, 50009 Zaragoza, Spain}
\affiliation{Max-Planck-Institut f\"ur Physik (Werner-Heisenberg-Institut), F\"ohringer Ring 6, 80805 M\"unchen, Germany}

\author{Yvonne Y. Y. Wong}
\affiliation{School of Physics, The University of New South Wales, NSW 2052 Kensington, Sydney, Australia}

\preprint{CPPC-2022-14}

\date{\today}
\smallskip
\begin{abstract}
In the scenario in which QCD axion dark matter is produced after inflation, the Universe is populated by large inhomogeneities on very small scales. Eventually, these fluctuations will collapse gravitationally to form dense axion miniclusters that trap up to $\sim$75\% of the dark matter within asteroid-mass clumps. Axion miniclusters are physically tiny however, so haloscope experiments searching for axions directly on Earth are much more likely to be probing ``minivoids''---the space in between miniclusters. This scenario seems like it ought to spell doom for haloscopes, but while these minivoids might be underdense, they are not totally devoid of axions. Using Schr\"odinger-Poisson and N-body simulations to evolve from realistic initial field configurations, we quantify the extent to which the local ambient dark matter density is suppressed in the post-inflationary scenario. We find that a typical experimental measurement will sample an axion density that is only around 10\% of the expected galactic dark matter density. Our results are taken as \emph{conservative} estimates and have implications for experimental campaigns lasting longer than a few years, as well as broadband haloscopes that have sensitivity to transient signatures. We show that for a $\mathcal{O}$(year)-long integration times, the measured dark matter density should be expected to vary by 20--30\%.
\end{abstract}

\maketitle

\section{Introduction}
\label{sec:introduction}
Axions have rapidly been accumulating interest in both the theoretical~\cite{DiLuzio:2020wdo, Chadha-Day:2021szb} and experimental~\cite{Irastorza:2018dyq, Semertzidis:2021rxs, Adams:2022pbo} particle physics communities recently. A particularly well-motivated subset of axion models is the quantum-chromodynamics (QCD) axion of Peccei \& Quinn~\cite{Peccei:1977hh, Peccei:1977ur}, which solves the puzzle of the missing neutron electric dipole moment~\cite{Peccei:1977hh, Peccei:1977ur, Weinberg:1977ma, Wilczek:1977pj, Kim:2008hd, Kim:1979if, Shifman:1979if, Dine:1981rt, Zhitnitsky:1980tq}. Simple arguments show that the QCD axion is also a rather minimal explanation for the abundance of dark matter in the Universe~\cite{Abbott:1982af, Dine:1982ah, PhysRevLett.48.1156, Wantz:2009it}. As a result, a flourishing campaign of new experiments now seeks to test whether these particles really are what constitute the invisible halos that envelop galaxies like our own Milky Way. 

Much of the recent interest in the QCD axion is driven by the upcoming generation of terrestrial direct detection experiments known as ``haloscopes’’~\cite{Sikivie:1983ip} which aim to explore the large swathes of open parameter space in the coming decades~\cite{Adams:2022pbo}. A diverse array of experimental techniques have been proposed, covering multiple interaction channels of the axion with the Standard Model, including the axion's defining coupling to the gluon~\cite{JacksonKimball:2017elr, Berlin:2022mia, Arvanitaki:2021wjk, Kim:2022ype}, as well as model-dependent couplings to photons~\cite{Salemi:2021gck, Aja:2022csb, Schutte-Engel:2021bqm, Lawson:2019brd, Millar:2022peq, TheMADMAXWorkingGroup:2016hpc, Beurthey:2020yuq, Berlin:2020vrk, BREAD:2021tpx, Brouwer:2022bwo, DMRadio:2022pkf, Oshima:2021irp, Baryakhtar:2018doz, Gramolin:2020ict, Arza:2021ekq, Zhang:2021bpa}, and fermions~\cite{Crescini:2016lwj, Ikeda:2021mlv, Mitridate:2020kly, Chigusa:2020gfs, OHare:2020wah, Graham:2020kai, Bloch:2019lcy, Garcon:2019inh}. 

The primary challenge in axion direct detection is both one of \textit{breadth}---the available mass range spans potentially $m_a\in [\sim 10^{-11},\,\mathcal{O}(1)]\,\mathrm{eV}$~\cite{Baryakhtar:2020gao, DEramo:2022nvb,Notari:2022zxo}---but also \textit{depth} in that the couplings to Standard Model particles may be vanishingly small if the axion was produced at a very high energy scale. Therefore, the field requires both resonance-based experiments that can search deeply while accessing tiny couplings, but also broadband-sensitive experiments that can search with lower sensitivity over wider mass windows.

Since the parameter space is still wide open, it is useful to think about possible theoretical biases we might have towards one axion model over another. In this regard, a target that is certainly worthy of investigation is the $10^{-5}$—$10^{-3}$~eV axion mass window predicted under the so-called ``post-inflationary’’ scenario. This is the scenario in which the Peccei-Quinn (PQ) symmetry---of which the axion is the associated Goldstone boson---is broken after the end of inflation. The resulting dark matter abundance turns out to be tied precisely to the energy scale of PQ breaking, and thus to the axion mass. Because the symmetry breaking occurs after inflation, large field gradients and topological defects emerge in our Universe. These highly non-linear features are generally not amenable to any analytical treatment, so dedicated cosmological field simulations are required to establish the precise relationship between the axion model parameters and the resulting dark matter abundance. To this end, the first simulation-backed predictions of the QCD axion abundance and mass have been made in recent years~\cite{Fleury:2015aca, Klaer:2017ond, Vaquero:2018tib, Gorghetto:2018myk, Buschmann:2019icd, Gorghetto:2020qws, Buschmann:2021sdq, OHare:2021zrq, Hoof:2021jft}, although many unresolved theoretical issues remain.

\begin{figure*}
    \centering
    \includegraphics[width=0.93\textwidth]{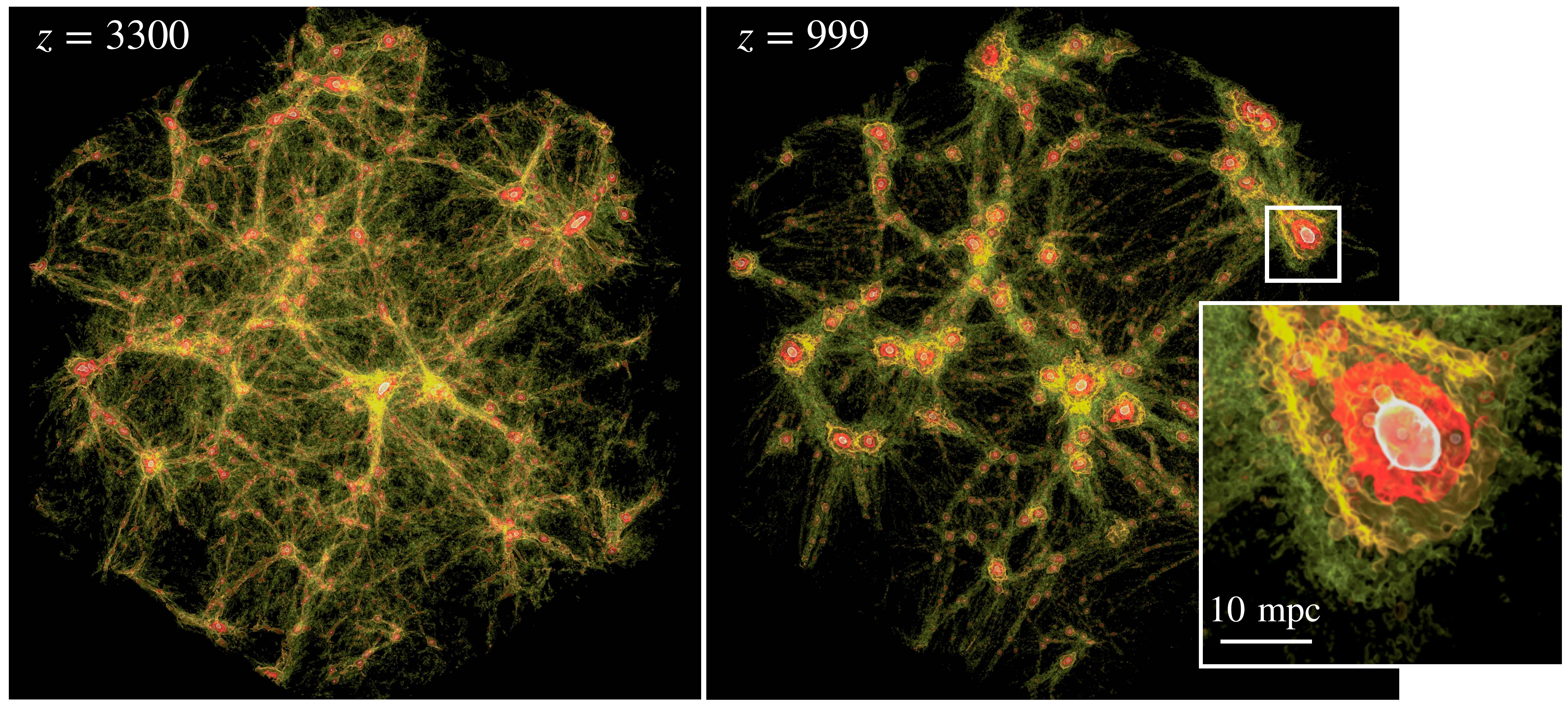}
    \caption{Volume rendering of the axion energy density at redshifts $z=3300$ (left) and $z=999$ (right) obtained from the N-body simulation in a $L = 8L_1 = 0.2\,\mathrm{pc}$ box. The colour scale denotes the logarithm of the overdensity field in a $512^3$ grid built from particle positions in the simulation. A  closeup of a particular minicluster halo indicates the physical scale relevant to the problem considered here. For reference, the Solar System moves a distance of about 0.2~mpc in one year.}
    \label{fig:MainCover1}
\end{figure*}

Independently of the precise value of the axion mass, one generic consequence of the post-inflationary scenario that has been known for some time is the production of dark matter substructures in a galactic halo~\cite{Hogan:1988mp, Kolb:1994fi, Kolb:1995bu, Zurek:2006sy, Kolb:1994fi, Hardy:2016mns, Davidson:2016uok, Enander:2017ogx, Blinov:2019jqc}. The large-amplitude yet ultra-small-scale density perturbations left over by the axion field dynamics around the QCD era lead to the formation of gravitationally-bound axion clumps already during radiation-domination. Previous studies have shown that around 75\% of all axions become bound into these so-called ``miniclusters’’~\cite{Eggemeier:2019khm}, which have masses similar to those of asteroids, i.e., \mbox{$M \in [10^{-13},10^{-9}]\,M_\odot$}, and radii of the order of an astronomical unit~\mbox{(AU$\sim 5\,\upmu$pc)}. In contrast with the average dark matter density in the solar neighbourhood 
(\mbox{$\rho_{\rm DM} \sim 0.01\, M_\odot$~pc$^{-3} \approx 0.4\,{\rm GeV}\,{\rm cm}^{-3}$}) inferred from stellar dynamics on scales $\gtrsim 100$~pc~\cite{Read:2014qva, Evans:2018bqy, deSalas:2020hbh}, these miniclusters are orders of magnitude denser than the average expectation.  But this statement also implies that the vast majority of the space \textit{outside} of miniclusters must  be relatively empty.

Importantly, axion miniclusters remain stable during the formation of the first galaxy-sized halos and are only tidally disrupted much later by close encounters with comparatively point-like objects like stars~\cite{Tinyakov:2015cgg, Dokuchaev, Kavanagh:2020gcy, Shen:2022ltx, Dandoy:2022prp}. Thus, as a firm \emph{prediction} of the post-inflationary axion dark matter scenario, the late-time consequences of axion miniclusters cannot be ignored in the context of direct or indirect searches. 

The existence of significant axion dark matter substructures in galaxy-sized halos has both positive and negative implications for efforts towards its detection. On the one hand, the presence of small bound clumps potentially opens up new opportunities for \textit{indirect} detection, such as the search for transient radio signals from axion miniclusters colliding with the intense magnetospheres of neutron stars~\cite{Tkachev:2014dpa, Edwards:2020afl}. Another proposed indirect detection technique is gravitational microlensing~\cite{Kolb:1995bu, Fairbairn:2017dmf, Fairbairn:2017sil}, although axion miniclusters might be too diffuse to generate a detectable signal~\cite{Ellis:2022grh}.%
\footnote{This result had been obtained for a subset of miniclusters that can be well described by a Navarro-Frenk-White profile~\cite{Navarro:1995iw}. For the majority of simulated miniclusters, however, the characteristic scale radius could not be resolved, and simulations of higher spatial resolution are required for a more in-depth analysis.}
On the other hand, the chances of a \textit{direct} detection of axion dark matter are significantly reduced if most of the axions are bound in miniclusters. A simple back-of-the-envelope calculation of the encounter rate of an experiment in a galaxy full of miniclusters suggests that we could only expect to be lucky enough to pass through one such substructure once every 100,000~years~\cite{Irastorza:2018dyq}.

While the later tidal disruption of miniclusters may slightly refill the space inside the galaxy with axions and facilitate their detection~\cite{Knirck:2018knd, OHare:2017yze}, we cannot eliminate the possibility that our experiments sit in a ``minivoid'' well away from centres of the bound axion clumps.  In such a worst-case scenario, it is quite possible that there are simply not enough axions in the space we occupy for terrestrial haloscope searches to work as intended.  Our aim in this work, therefore, is to quantify how bad it can get: how much of the space in a galaxy has a deficit of axion dark matter and how big this deficit is.  To do so, we shall perform simulations of the axion minicluster formation and evolution. We begin from realistic initial conditions left over from the PQ and QCD phase transitions that account for the decay of topological defects, and the formation and collapse of temporary soliton-like field configurations called axitons~\cite{Vaquero:2018tib, Buschmann:2019icd} that lay down the fluctuations from which the miniclusters form. Our simulations then follow the gravitational dynamics of the density fields across matter-radiation equality.

The formation and evolution of axion miniclusters have previously been studied using semi-analytic techniques such as the Press-Schechter~\cite{Fairbairn:2017sil, Enander:2017ogx, Blinov:2019jqc} and the Peak-Patch~\cite{Ellis:2020gtq, Ellis:2022grh} formalisms, as well as in full numerical N-body simulations~\cite{Eggemeier:2019khm, Xiao:2021nkb, Shen:2022ltx}. Our work extends upon and complements the N-body studies of~\cite{Eggemeier:2019khm, Xiao:2021nkb, Shen:2022ltx}, and uses a mixture of Schr\"odinger-Poisson and N-body methods to simulate the formation, growth, and merger of halos made of miniclusters. We furthermore quantify our results not only in terms of the properties of the axion miniclusters, but also the statistics of the axions that are \textit{not} gravitationally bound in substructures. A visualisation of the axion density field produced in our simulations at two times during the evolution is shown in Fig.~\ref{fig:MainCover1}.

The rest of the article is organised chronologically in cosmic time. We begin in Sec.~\ref{sec:earlyuniverse} with a description of the results of the early-universe lattice simulations that provide us with the requisite initial conditions. We switch to simulations of the axion's gravitational dynamics in Sec.~\ref{sec:gravity} and study the properties of the resulting axion minivoids in Sec.~\ref{sec:voids}. We discuss the consequences of these results for haloscopes in Sec.~\ref{sec:haloscopes} and, finally, conclude in Sec.~\ref{sec:conc}.

\section{Early-Universe simulations}\label{sec:earlyuniverse}

\subsection{Minicluster seeds}

The initial conditions are of critical importance when studying the later formation of miniclusters, and generating a suitable initial configuration for the field requires simulating nonlinear dynamics of the axion well before gravitational effects become relevant. In the post-inflationary scenario, the dynamics of the axion field goes through several distinct eras during which important and nontrivial field configurations emerge. Firstly, global axion strings form as a result of the $U(1)_{\rm PQ}$ spontaneous symmetry breaking at very large temperatures $T\gtrsim 10^{10}$~GeV. Axion strings then enter a scaling regime and survive until around the QCD phase transition when the axion mass becomes comparable to the Hubble parameter, $H(t)$. At this point, the axion field begins damped oscillations around the minimum of its potential and domain walls form between strings. We call the redshift or time when this happens the \emph{characteristic time}, $t_1$ or $z_1$, and can be thought of as a measure of when in cosmic evolution the axion becomes dark matter in earnest. 

The characteristic times can be calculated by satisfying the condition $m_a(t_1)=H(t_1)$. This redshift turns out to be very high, and depends weakly on the axion mass
\begin{equation}
     z_1=2.5\times 10^{13}\left(\frac{0.5~{\rm meV}}{m_a}\right)^{\frac{2}{n+4}} \, .
\end{equation} 
The comoving size of the causal volume at $t_1$ is, accordingly, the characteristic \textit{length scale}\footnote{Note that our choice $m_a=0.5$ meV, motivated by Ref.~\cite{Ellis:2022grh}, leads to a slightly different definition of $L_1$ with respect to Ref.~\cite{Eggemeier:2019khm}.},
\begin{equation}
    L_1=0.025~{\rm pc}\left(\frac{0.5~{\rm meV}}{m_a}\right)^{\frac{2}{n+4}},
\end{equation} 
where $n$ is a parameter that controls the growth of the axion mass with temperature, i.e., $m^2_a \sim T^{-n}$. Assuming $n\simeq 7$~(see, e.g., \cite{Borsanyi:2016ksw}), we obtain $L_1\sim m_a^{-0.18}$. Therefore, given the fact that $m_a$ cannot be arbitrarily varied over many orders of magnitude---as there is only a restrictive mass range that give the correct dark matter abundance---the spatial scale $L_1$ falls between 0.02~pc and 0.05~pc, i.e., it does not vary substantially. 

After a few multiples of $t_1$, the string-wall system collapses under the wall tension,%
\footnote{An unstable network is required, in order to avoid the domain wall problem. This is satisfied in models with $N_{\rm DW}=1$ (e.g., KSVZ-like) or by introducing a bias term~\cite{Barr:1982uj, Lazarides:1982tw, Dvali:1994wv, Chang:1998bq, Rai:1992xw, Reig:2019vqh, Caputo:2019wsd, Gelmini:1988sf,Gelmini:2020bqg,Gelmini:2021yzu,Gelmini:2022nim} (however see also Ref.~\cite{Beyer:2022ywc}).} during which almost the entirety of the energy of the system is converted into nonrelativistic modes of the axion field. The field can then be treated as an inhomogeneous nonrelativistic fluid.
Except for a few $r\sim m^{-1}_a$ localised regions which form quasi-stable solitonic configurations of the axion field known as \emph{axitons}, the field is largely in the linear regime.  Axitons are expected to disappear when the axion mass reaches its present-day value~\cite{Buschmann:2019icd}.

The topological defects and the subsequent axitons both leave behind large overdensities in the energy density distribution that act as \emph{seeds} for gravitational structures to form later in the cosmological evolution. In our recent work \cite{OHare:2021zrq} we performed high-resolution simulations of this early stage as a function of the axion mass growth index, $n$, and studied the resulting distribution of minicluster seeds. We continue the evolution of this system here, when the field values are small enough that the nonrelativistic approximation is safe to assume. 

For concreteness, we focus only on the QCD axion, adopting the value $n=7$. Our initial configurations are the final snapshots of two previous simulations with $4096^3$ and $8192^3$ lattice sites (see Appendix \ref{app:methods} for further details), and box side lengths of $L=8L_1$ and $L=16L_1$ respectively.\footnote{Going in, we know that these results, and by extension the initial structure of our miniclusters seeds, will suffer from two main uncertainties plaguing all simulations of this type: (i) the smallness of the simulated string tension compared to its physical value at $t\sim t_1$ (see, e.g., \cite{Gorghetto:2020qws, Klaer:2017ond}), and (ii) the effects of small-scale structure to the energy density distribution left by axitons that cannot be resolved until their disappearance by current simulations \cite{Vaquero:2018tib}. These issues are the subject of ongoing investigations.}

\subsection{The Schr\"odinger-Poisson system}

In the linear regime, the dynamics of the axion field can be described as the slow free-streaming of almost-frozen relic waves.
Gravitational effects, on the other hand, can be accounted for in the weak-field limit, where it is convenient to work in the Newtonian gauge defined by the line element ${\rm d} s^2=-(1+2\Phi_N) {\rm d}t^2 + R^2(t) (1-2\Phi_N) {\rm d} x^i {\rm d} x_i$, where $R(t)$ is the scale factor in a Friedmann-Lema\^itre-Robertson-Walker background, and $\Phi_N \ll 1$ can be identified with the Newtonian gravitational potential.

Given that the now nonrelativistic axion field is rapidly oscillating with its mass $m_a$  as the leading order frequency, we can employ a WKB approximation to write the axion field $a(\mathbf{x})$ as
\begin{equation}
    a(\mathbf{x})=\left(\frac{1}{\sqrt{2m_a}}\psi e^{-im_at }+\text{h.c.}\right),\label{eq:paxion}
\end{equation} 
where $\psi$ denotes a slowly varying complex scalar field. 
In the limit  $\partial_t\psi\ll m_a\psi$,
the evolution of $\psi$ is governed by the Schr\"odinger-Poisson equations,
\begin{align}
    i\partial_{t}\psi&=-\frac{\nabla^2\psi}{2m_{a}}+m_{a}\Phi_N\psi,\label{eq:sp1} \\
    \nabla^2\Phi_N&=\frac{4\pi G}{R}\delta_a\vert\langle\psi\rangle\vert^2,\label{eq:sp2} 
\end{align} 
where $\delta_a \equiv \rho_a/\langle\rho_a\rangle-1$ is the axion overdensity, $\rho_a=m_a\vert\psi\vert^2$ is the energy density, and $\langle \cdots \rangle$ denotes an average over the simulation volume.

The set of equations~\eqref{eq:sp1} and \eqref{eq:sp2} describes any scalar field in the classical regime, i.e., when the occupation number is very large. Their solutions exhibit wave-like effects such as interference patterns and the formation of soliton solutions, as have been observed in several numerical studies~\cite{Guzman2004, Schive2014_Nature, Schive2014_PRL, Schwabe2016, Veltmaat2018, Levkov2018, Eggemeier:2019jsu, Axionyx, Schwabe:2021jne, Eggemeier2021} in different cosmological scenarios. These features are expected to show up on scales smaller than the so-called ``Jeans wavelength''~\cite{Hu:2000ke}, i.e.,
\begin{equation}
    \lambda_{\rm J}=\frac{2\pi}{R(16\pi G\rho_am^2_a)^{1/4}}, \label{eq:laJ}
\end{equation} 
or the de Broglie wavelength $\lambda_{\rm dB}=2\pi/(m_av_a)$.
For axion masses $m_a\gtrsim 10^{-5}$ eV, $\lambda_{\rm J}$ and $\lambda_{\rm dB}$ are comparable to, or smaller than, the discretisation scale of a typical simulation  ($L=8L_1$ box size and $\sim 4000^3$ points; see, e.g., Fig.~\ref{fig:lambdaJ} in Appendix~\ref{app:methods}).
In contrast, the Schr\"odinger-Vlasov correspondence~\cite{Uhlemann:2014npa} stipulates that the evolution of $\rho_a$ on scales \textit{larger} than $\lambda_{\rm J}$ cannot be distinguished from that of collisionless pressureless matter such as standard particle cold dark matter.
Therefore, standard N-body simulations can be used to evolve the axion field,
provided that wave-like effects appear only on scales smaller than a simulation's spatial resolution~\cite{Eggemeier:2019khm}.

We solve the system of equations~\eqref{eq:sp1} and \eqref{eq:sp2} beginning at an initial redshift of $z=5\times 10^{11}$, until the numerical solution of the Schr\"odinger-Poisson system becomes unreliable at $z\sim 10^6$ (discussed further in Appendix~\ref{app:methods}) and the first miniclusters start to form. From this point onwards, we need to switch to N-body simulations in order to capture the gravitational collapse of axion overdensities; this will be discussed in the next section. Because in addition to free-streaming, we include the effects of linear gravity between the early Universe simulations and the beginning of the N-body simulations in our modelling, this work represents an improvement upon the numerical study of minicluster formation of Ref.~\cite{Eggemeier:2019khm}. 
Additional details on the numerical evaluation can be found in Appendix \ref{app:methods}.

\section{N-body simulations}\label{sec:gravity}

The largest overdensities are expected to collapse and decouple from the Hubble flow to form gravitationally bound structures at redshifts $z\sim 10^6$. As briefly justified above, from this point onwards we switch to N-body simulations and continue the numerical evolution of the axion field under gravitational interactions using the N-body code~\texttt{GADGET-4}~\cite{Springel:2020plp}.  We use box sizes corresponding to $L/L_1=8,16$ and compare our simulation results with those of Ref.~\cite{Eggemeier:2019khm} that used a larger box size of $L/L_1=24$. Even though the initial conditions of the latter have been evaluated without the Schr\"odinger-Poisson evolution, we will show that our main results on the minivoid distribution and statistics are largely independent of the choice of initialisation method. An in-depth analysis of the choice of initial conditions and their implications on the structure of axion miniclusters will be discussed in a forthcoming publication.   Figure~\ref{fig:MainCover1} shows two visualisations of the resulting density field from our $L/L_1 = 8$ simulation.  

\subsection{Initial conditions}
\label{sec:initial}

In order to create the initial conditions for the N-body evolution, the axion field values on the lattice need to be converted into particle configurations. One way to do this, explored in Ref.~\cite{Eggemeier:2019khm}, is to create particles of the same mass value and vary the local number of particles placed in the simulation according to the local density contrast. 
Specifically, a number of particles equal to ${\rm floor}(\rho_i/\langle\rho\rangle)$ are created on every grid point $i$ and each particle is randomly displaced by an amount drawn from a Gaussian distribution with a standard deviation equal to a quarter of the grid spacing. This method is particularly efficient in describing the small-scale structure and the density profiles of the miniclusters, at the cost of a lower resolution in the initially underdense regions that eventually become minivoids. 

To better sample the underdense regions, we explore in this work an alternative initialisation method.  Here, we place particles in the initial snapshot \textit{homogeneously}, i.e., one particle per grid cell, but each particle can now carry a \textit{different mass} that reflects the axion density on the grid cell. Specifically, we scale masses according to $m_i=\delta_{a,i}m_{\rm av}$, where $m_{\rm av}$ is the average particle mass. In what follows, we refer to the first mapping method as the \enquote{same mass ICs} and the second method as the \enquote{different mass ICs}. 
As in Ref.~\cite{Eggemeier:2019khm}, our lattice grids must be down-sampled because of limitations on the total particle number tractable by available computational resources; currently, we can simulate up to $1024^3$ particles, which shrinks the lattice grids by a factor of 4--8 per spatial dimension when switching to the N-body method.

We note in passing that we have also implemented another refinement in the initialisation procedure by including initial particle \textit{velocities}, which had been neglected in Ref.~\cite{Eggemeier:2019khm}.  To do this, we computed the gradient of the phase of $\psi$, i.e.,~$v_i=\nabla \arg\psi_i/m_a$, for the $i$th particle. 
However, in short, we did not observe many substantial differences between including and not including the velocity information---at least not in the context of the minivoid analysis we shall present shortly.

%%%%%%%%

\subsection{Final simulation time}
Having implemented the initial particle realisation, their subsequent evolution under gravitational interaction as tracked by \texttt{GADGET-4} simply follows well-established and widely used N-body techniques. The only remaining aspect that needs to be discussed is the final time imposed upon the simulation by the finite box size.

Because we use boxes of several different sizes, the final simulation redshifts also vary slightly. In the $L=24L_1$ simulation box of Ref.~\cite{Eggemeier:2019khm}, the N-body simulation becomes unreliable past $z=99$. This can be seen from the dimensionless power spectrum $\Delta^2(z)$, which becomes $\mathcal{O}(1)$ on the largest simulated scales $k\sim 1/L$ at these late times. The simulated power spectrum on these large scales also begins to deviate from the linearly-evolved power spectrum, given by $\Delta^2(k,z_i)D^2(z)$,
where
\begin{equation}
    D(z)=1+\frac{3}{2}\frac{1+z_{\rm eq}}{1+z}\,
\end{equation} 
is the linear growth factor, $z_i$ the initial redshift, and $z_{\rm eq}$ is the redshift at matter-radiation equality.  Note also that for the smaller simulation box sizes, $L/L_1=8,16$, used in this work, the final simulation redshift must necessarily be higher, because the largest simulated scales in these boxes are smaller and hence become nonlinear at earlier times. 

Here, we choose the final simulation redshift $z_f$ by demanding that the 
 ratio between the simulated power spectrum $\Delta^2(z_f)$ and the linearly-evolved one at \mbox{$k\lesssim 15~{\rm pc}^{-1}$} should not exceed some threshold $p_{\rm thr}$, i.e., 
\begin{equation}
    \left.\frac{\Delta^2(k,z_f)}{\Delta^2(k,z_i)D^2(z_f)}\right|_{k\lesssim 15~{\rm pc}^{-1}} < p_{\rm thr}.
\end{equation} 
The choice of $p_{\rm thr}=0.2$ gives us the following $z_f$ values:
\begin{align}
    z_f=999, &~~~~~~~~L=8L_1~=0.2~{\rm pc}\\
    z_f=499, &~~~~~~~~L=16L_1=0.4~{\rm pc}.
\end{align}

If we were to evolve past $z_f$, some small-scale properties such as the shapes of the miniclusters would unlikely be affected. However, statistical statements about the large-scale properties of the system would not be trustworthy. Given that we are quantifying the voids which take up the bulk of the simulation volume, the final redshift is something we must enforce. In the following, we will refer to the box size with its length in units of parsecs. Further details on the N-body simulations can be found in Appendix~\ref{app:methods}.

\section{Minivoids}\label{sec:voids}
While most of the mass of dark matter is contained in miniclusters and minihalos, these gravitationally bound objects make up only a tiny fraction of the total \textit{volume} of the simulation. The ``minivoids'' that span the space between the miniclusters are where we are mostly likely to be. Thus,  as argued in Sec.~\ref{sec:introduction}, it is of critical importance for direct detection that we quantify the energy density of axions in the voids. To our knowledge, an extensive study on void statistics for the axion post-inflationary scenario has not previously appeared in the literature. 

\subsection{Bound fraction}

Before we get to the minivoids, let us briefly remark on where the majority of axions actually find themselves. After matter-radiation equality, the evolution of the density field essentially consists of mergers of miniclusters, of typical mass \begin{equation}
    M_1\simeq 2\times 10^{-12}~M_{\odot}\left(\frac{0.5~{\rm meV}}{m_a}\right)^{\frac{6}{n+4}},
\end{equation} into minicluster halos with mass $M_{\rm MCH}\gg M_1$. Again, the impact of the axion mass is only marginal ($M_1\propto m_a^{-0.54}$), and the reference mass scale $M_1$ can only vary by roughly one order magnitude between $10^{-12} M_{\odot}$ and $10^{-11} M_{\odot}$. These halos are expected to contain most of the axions in the simulation, the extent of which can be quantified by calculating the \emph{bound fraction} $f_b$, defined as the ratio of the cumulative mass of gravitationally-bound axions in miniclusters and the total mass contained in the simulation volume.%
\footnote{More precisely, the gravitationally-bound axions correspond to groups of at least 32 particles identified by the \emph{friend-of-friends} halo finder algorithm in the simulation.} 

\begin{figure*}
    \centering
    \includegraphics[width=0.49\textwidth]{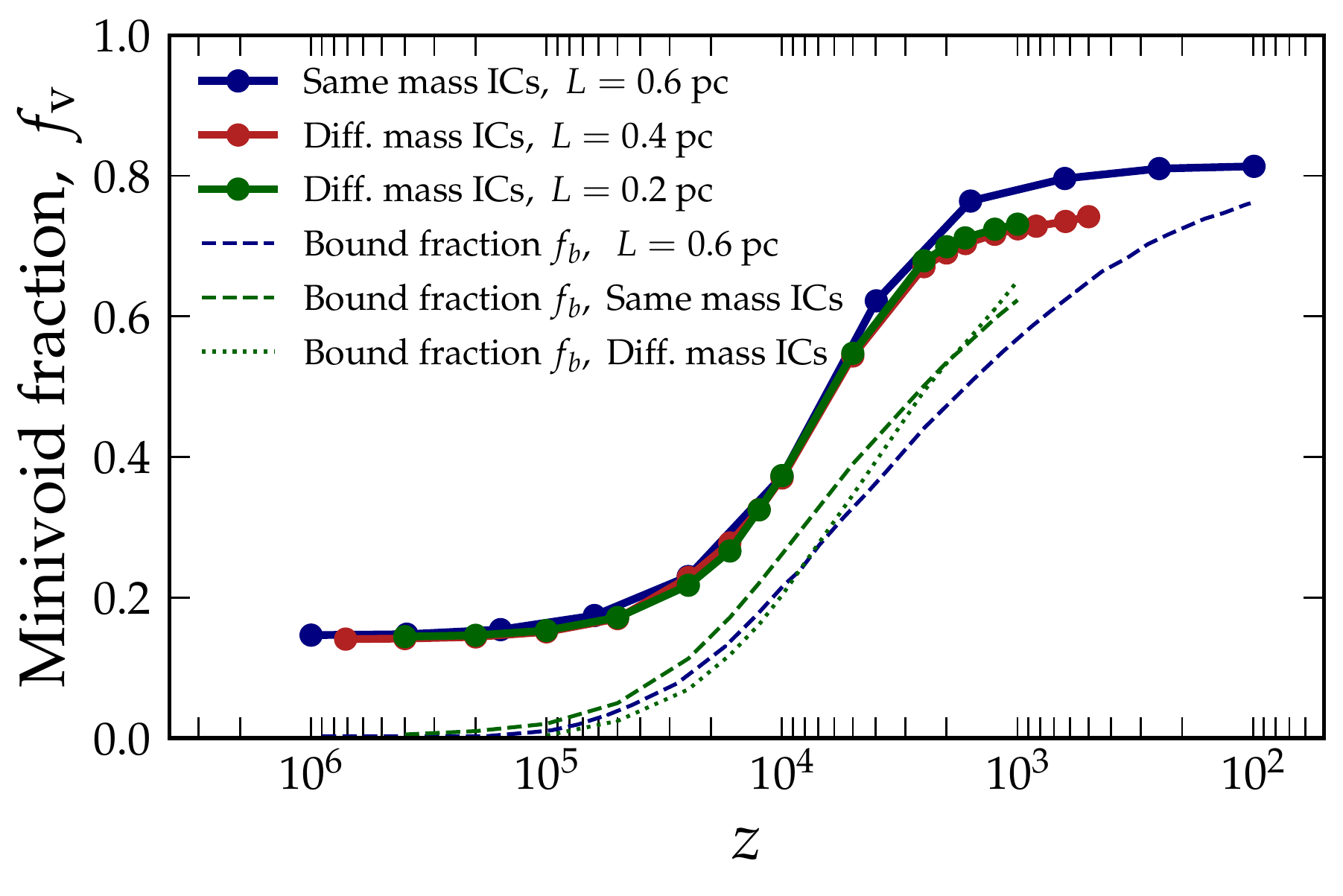}~
    \includegraphics[width=0.49\textwidth]{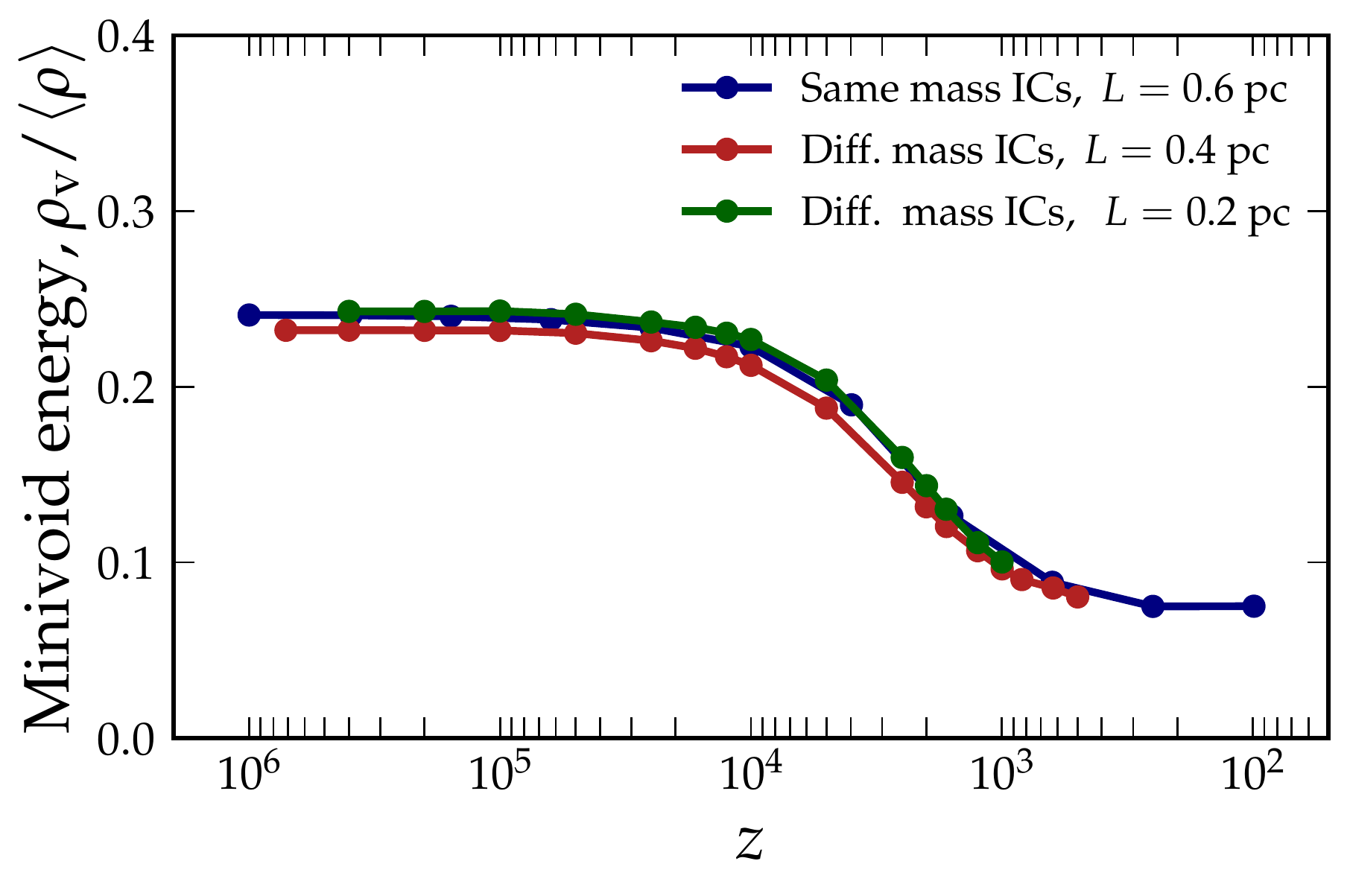}
    \caption{A comparison of void statistics between the different simulations using the \enquote{different mass ICs} method of this work and the \enquote{same mass ICs} method of Ref.~\cite{Eggemeier:2019khm} (see text for details). We adopt a density threshold value of $\delta^{\rm thr}_a=-0.7$ to identify voids in the simulation volume. \emph{Left}: Void volume fraction $f_{\rm v}$ (solids lines) and the minicluster bound fraction $f_b$ (dashed lines). \emph{Right}:  Average energy density in voids, $\rho_{\rm v}$, normalised to the average energy density over the whole simulation volume $\langle\rho\rangle$.}
    \label{fig:void}
\end{figure*}

The left panel of Fig.~\ref{fig:void} shows $f_b$ as a function of time as dashed and dotted lines. For the $L = 0.6$~pc simulation (dashed blue line), we recover the result of Ref.~\cite{Eggemeier:2019khm} that the fraction of axions bound in miniclusters reaches a value of $f_b\sim 0.75$ by around $z=99$ and begins to plateau.
Comparing this with the bound fraction in our $L=0.2$~pc simulations, where we distinguish between runs using the \enquote{same mass ICs} and the \enquote{different mass ICs} technique described in Sec~\ref{sec:initial},
it is evident from Fig.~\ref{fig:void} that the results across these different simulations are broadly similar.
In particular, at $z=999$ the bound fraction is $f_b\sim 0.6$, with a 5\% difference between the different simulations. This reassures us of consistency between the original ``same mass ICs'' initialisation procedure of Ref. \cite{Eggemeier:2019khm} and the ``different mass ICs'' approach proposed in this work.

\subsection{Finding minivoids}\label{sec:finder}
Given that $\sim$25\% of the axions are defined as being outside of miniclusters, 
we might already expect na\``{\i}vely that the typical density at a given point outside of miniclusters in the simulation to be $\sim 25$\% of the average density over the whole simulation volume.
However, this $\sim 25$\% suppression is likely to be an overestimate, as it assumes that the unbound axions are evenly distributed in the box, which is certainly not the case. We therefore analyse the structures that primarily fill the simulation volume directly to obtain a more precise quantification of the suppression of the average axion density.

We find void regions on the grid by first building the density contrast field from the particle snapshots. A Gaussian smoothing filter is then applied on the snapshot, using a range of filter radius between $R_{\rm min}=5\Delta_x$, where $\Delta_x$ denotes the mean particle separation distance (or the grid spatial unit $\Delta_x=L/N$), and $R_{\rm max}\sim L_1$. This choice of $R_{\rm min}$ ensures that our results are not contaminated by discretisation effects on small scales; on the other hand, for $R_{\rm max}$, we do not expect voids to have comoving radii larger than $\sim 1.5L_1\simeq 38$~mpc.
Void cells are tagged if their average density contrast is smaller than a pre-determined threshold $\delta^{\rm thr}_a$, and minivoids are identified as spherical regions that do not overlap with other minivoids previously found. 
We test thresholds in the range $\delta_a^{\rm thr}\in [-0.4,-0.75]$, with $\delta_a^{\rm thr}=-0.7$ as the fiducial value.%
\footnote{See, e.g., \href{https://github.com/franciscovillaescusa/Pylians3}{https://github.com/franciscovillaescusa/Pylians3}~\cite{Pylians}.} 
 We catalogue the voids by their radii, in order to study their density profiles and statistical distributions. 

As depicted in Fig.~\ref{fig:void}, we estimate the fraction $f_{\rm v}$ in volume occupied by minivoids, as well as their typical energy density $\rho_{\rm v}$, by gathering all the minivoids found in the previous steps. One might think that both $f_{\rm v}$ and $\rho_{\rm v}$ depend on the choice of density threshold $\delta_a^{\rm thr}$. This is so at the beginning of the simulation, but the dependence on our choice of $\delta_a^{\rm thr}$ clearly dies away after matter-radiation equality and towards the end of our simulations; see Fig.~\ref{fig:void8} in Appendix \ref{app:void}. In addition, our estimates of $f_{\rm v}$ differ from those of Ref.~\cite{Eggemeier:2019khm} by less than 5\% differences, and the minivoid energy density estimates are in excellent (sub-percent) agreement.
In particular, we noticed in the $L=0.2\,\mathrm{pc}$ simulation that these results are completely independent of the initialisation mapping method and whether or not particle velocities have been incorporated in the initial conditions. 
We find convergence to the following values: 
\begin{align}
    &\rho_{\rm v}/\langle\rho\rangle\simeq 0.080, ~~~f_{\rm v}\simeq 0.75, ~~~(z= 499),\\
    &\rho_{\rm v}/\langle\rho\rangle\simeq 0.075, ~~~f_{\rm v}\simeq 0.81, ~~~(z= 99). \label{eq:finalvoid}~~~
\end{align} 
Of the remaining $\sim 20\%$ of the total volume, miniclusters occupy only up to 1\%; the rest is in the form of slightly underdense or average-density regions ($\rho\sim\langle\rho\rangle$).  More details can be found in Appendix \ref{app:void}.

\subsection{Minivoid structure}

\begin{figure}
    \centering
    \includegraphics[width=0.48\textwidth]{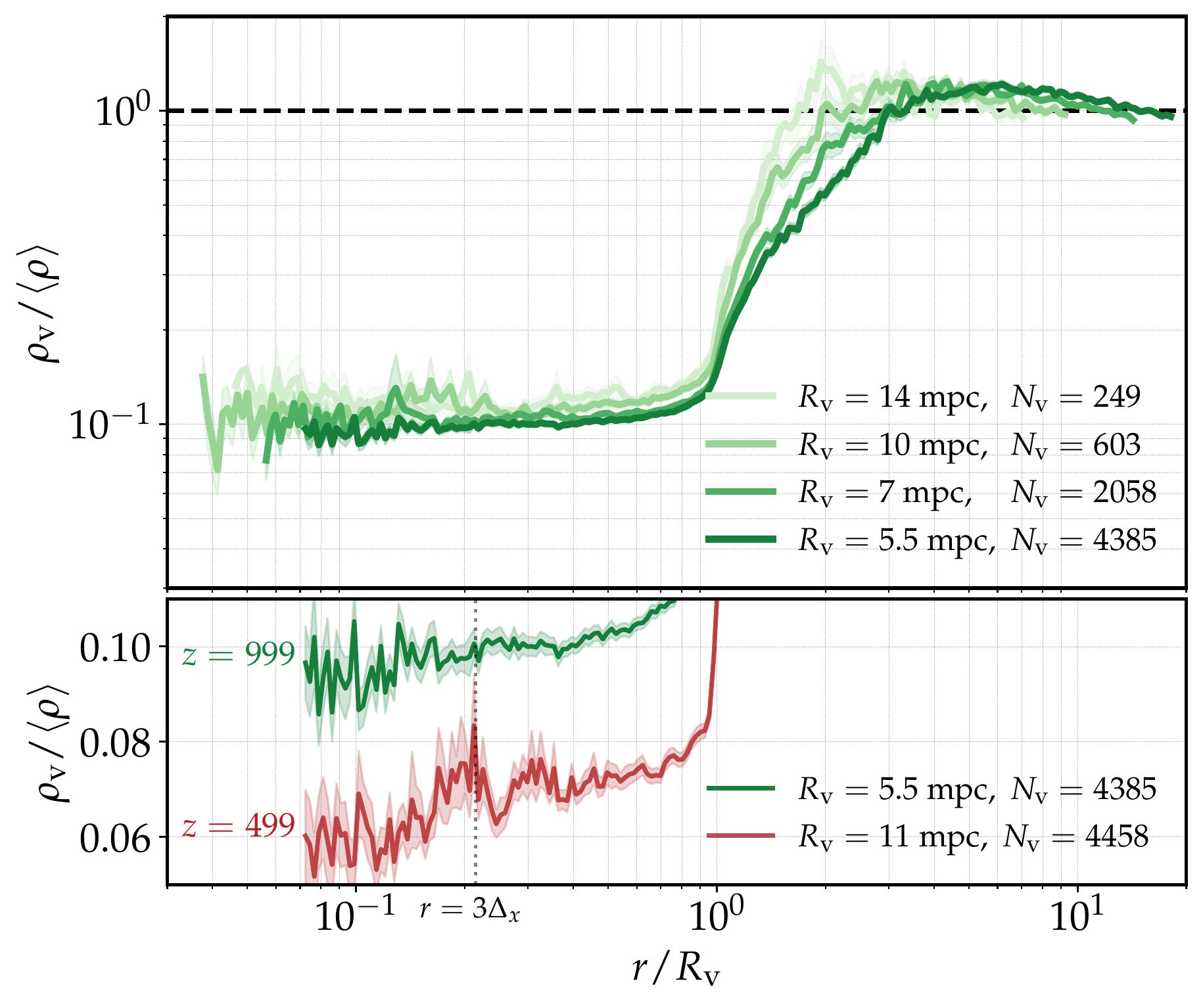}
    \caption{Minivoid density profiles as a function of the normalised radius. \emph{Top}: Profiles of minivoids of several different sizes found in the $L=0.2$ pc simulations at final time $z=999$.  \emph{Bottom}: Comparison between profiles obtained from the $L=0.2$ pc and $L=0.4$ pc simulations  at their respective final redshifts, $z_f=999$ and $z_f=499$. In both cases over $4000$ voids with $R_{\rm v}=14\Delta_x$ have been analysed. Shaded areas correspond to statistical errors over the $N_{\rm v}$ voids found.}
    \label{fig:vprof}
\end{figure}

Now we turn to the internal structure of voids by computing their density profiles.  The profiles are constructed by integrating over spherical shells around the centre of all voids of a given size, from a minimum radius $r_{\rm min}=\Delta_x$ to the maximum radius $r_{\rm max}=10R_{\rm v}$, where $R_{\rm v}$ denotes the void radius.
The upper panel of Fig.~\ref{fig:vprof} shows the void density profiles of all  voids with comoving radii $R_{\rm v}=5.5,\,7,\,10$, and 14~mpc identified in the $L=0.2$~pc simulation at redshift $z=999$.
We find an approximately constant profile within the void radius, and a slow decrease at the discretisation level, $r\lesssim 3\Delta_x$.  Immediately beyond $r\sim R_{\rm v}$, the void density increases steeply and overshoots the average value $\langle \rho \rangle$, before dropping back down to the average again at large radii well outside the void. 

We also compare in the lower panel of Fig.~\ref{fig:vprof} the density profiles of voids with the smallest radius, $R_{\rm v}=5.5$~mpc, from the $L=0.2$~pc simulation at redshift $z=999$, to those of a comparable population from the  $L=0.4\,\mathrm{pc}$ simulation at $z=499$, which have $R_{\rm v}=11$~mpc. This close-up reveals spiked regions in the void profiles at radii $r<R_{\rm v}$ that are even more pronounced for larger~$R_{\rm v}$. We attribute these spikes to small miniclusters that have formed at the earliest times but have not merged into larger structures. 
Notwithstanding the highly nontrivial substructure on $\sim$mpc scales, the average density profiles of the minivoids---which incorporates full particle information from the simulation---agree with the result obtained earlier from the void finding procedure, which uses smoothed energy densities on the grid. This is seen by comparing the right panel of Fig.~\ref{fig:void} with the bottom panel of Fig.~\ref{fig:vprof} at radii $r\gtrsim 3\Delta_x$, where $\rho_{\rm v}/\langle\rho\rangle\simeq 0.1$ at $z=999$ and $\rho_{\rm v}/\langle\rho\rangle\simeq 0.075$ at $z\lesssim 499$.

\begin{figure}
    \centering
    \includegraphics[width=0.48\textwidth]{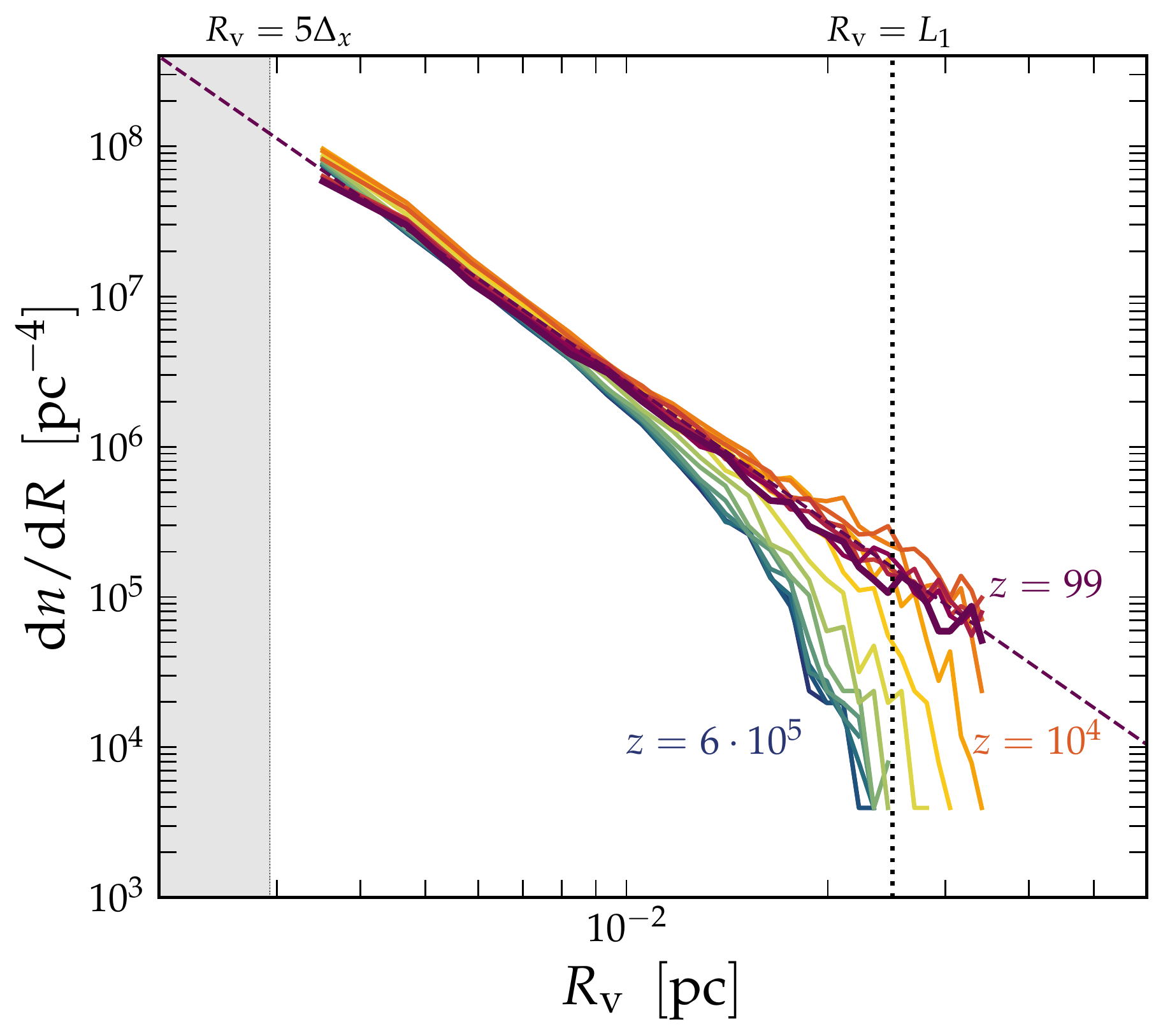}
    \caption{Minivoid size function at various redshifts extracted from the $L=0.6$~pc simulation. After matter-radiation equality, the distribution of voids follows a $R^{-3}$ power law between comoving radii $R_{\rm v}\sim 3$ mpc and $R_{\rm v}\sim 35$ mpc.}
    \label{fig:vsf}
\end{figure}

Similarly to the halo mass function for the miniclusters, we can describe the minivoid statistics by computing the \emph{minivoid size function}, defined as the comoving number density of minivoids per differential radius interval $\mathrm{d}n/\mathrm{d}R$. Figure~\ref{fig:vsf} shows the distribution of minivoids from the simulation with the largest statistics, namely, the one from Ref.~\cite{Eggemeier:2019khm}. We find that relatively large void regions start to form even before matter-radiation equality, and by $z\sim 1000$ the shape of the minivoid size function does not change anymore. 
Between $R_\mathrm{v} = 5\Delta_x$ and $R_\mathrm{v} = L_1$, the void size function is almost scale-invariant and can be described by a power law,
\begin{equation}
    \frac{{\rm d}n}{{\rm d}R}=a R^{-b}\,,
\end{equation} 
where the parameter values $a=1.43$ and $b=3.11$ provide a good fit to the simulation results at $z=99$, as indicated by the dashed line in Fig.~\ref{fig:vsf}. Note that our smaller  $L=0.2,\,0.4$~pc simulations exhibit the same void size functions and a comparable redshift dependence, indicating that 
these results are generally convergent.  See Appendix~\ref{app:void} for further details.

\subsection{Minivoid fluctuations}

\begin{figure*}[t]
    \centering
    \includegraphics[width=0.49\textwidth]{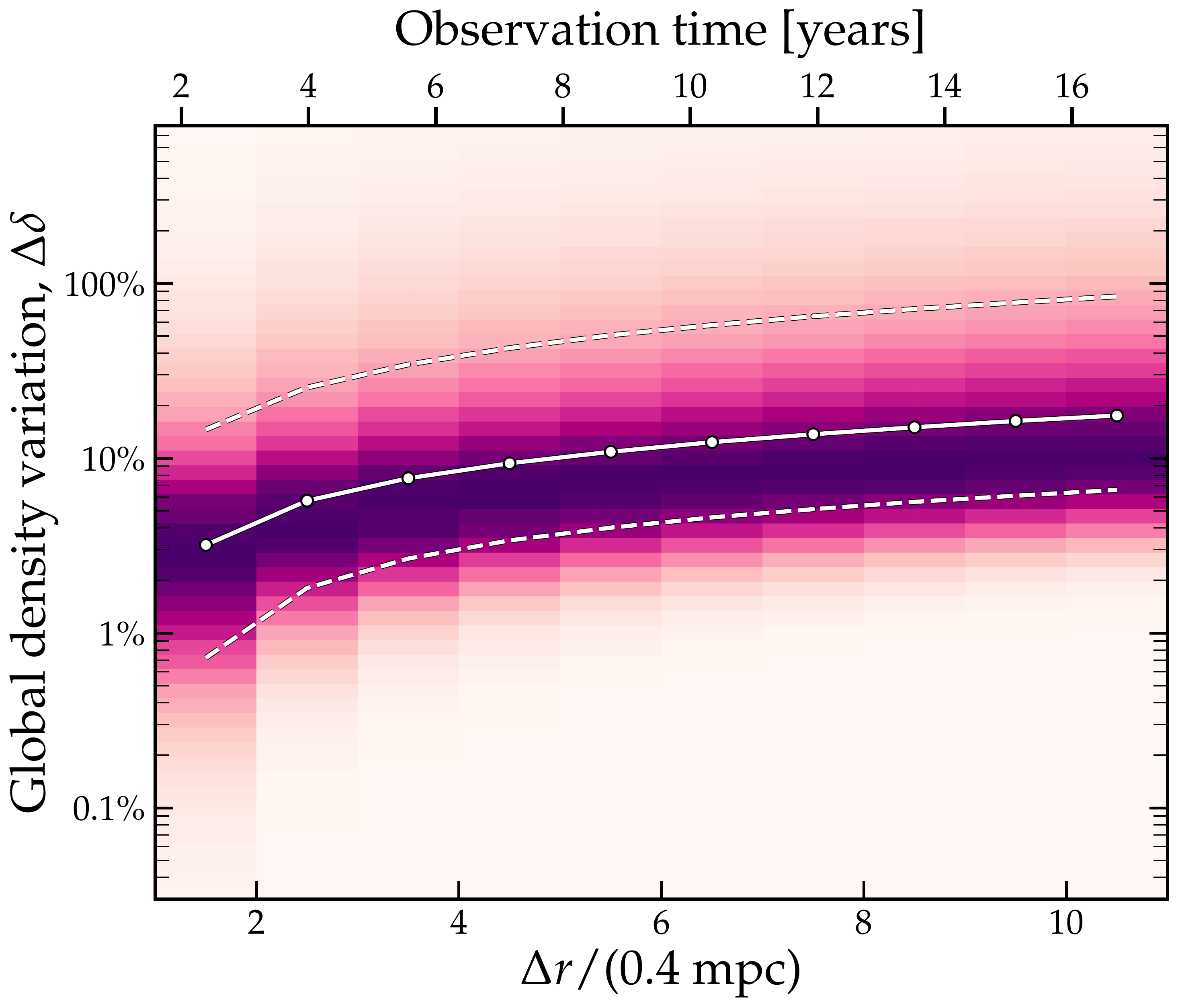}
    \includegraphics[width=0.49\textwidth]{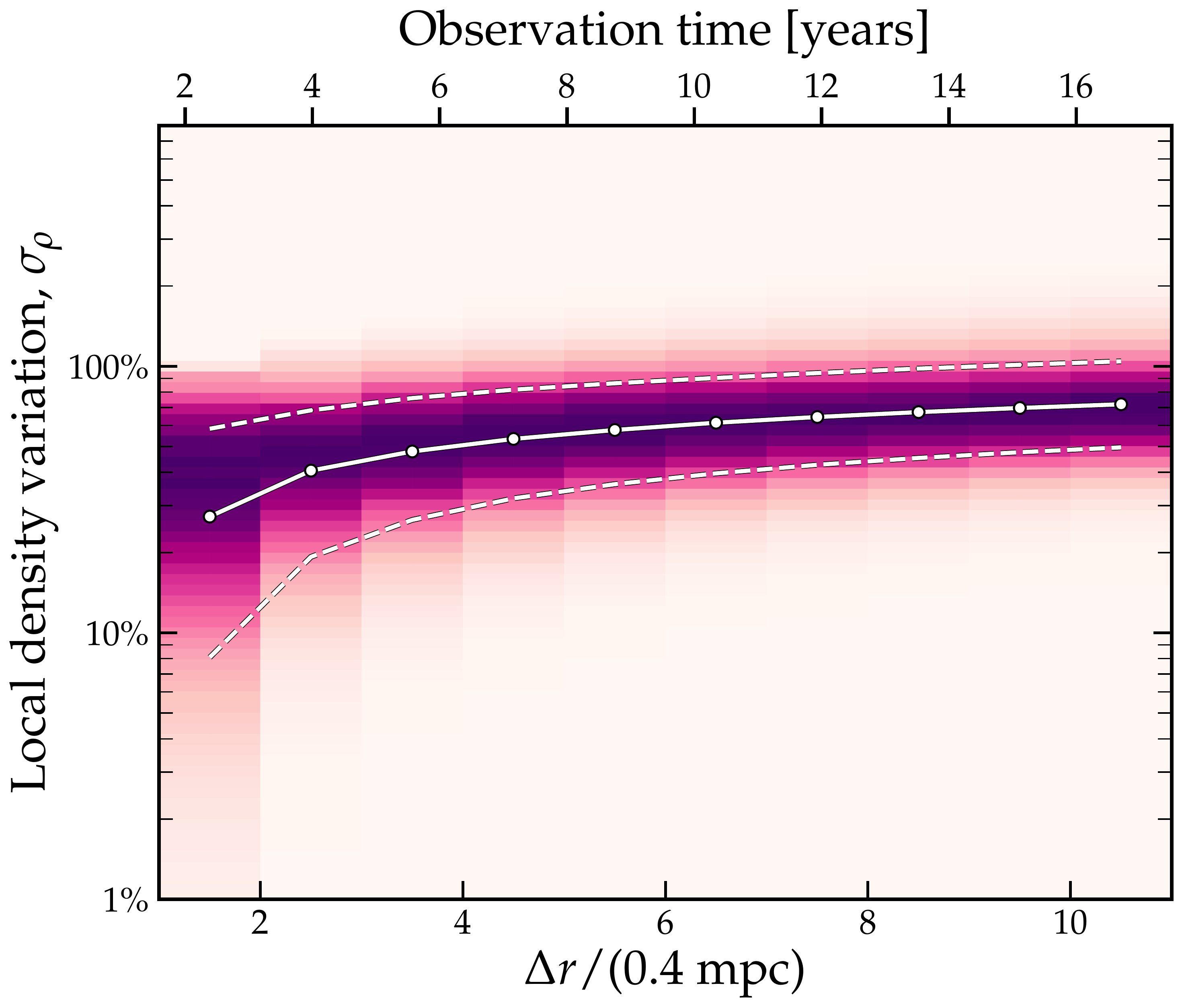}
    \caption{Two measures of the density variation along random trajectories through the simulation volume, quantified in terms of $\Delta r$, or, equivalently, the observation time, assuming the Solar System moves around the galaxy at a speed of 246 km/s. The left panel shows the \textit{global density variation}, $\Delta \delta$, defined in Eq.~\eqref{eq:deltadelta}, which describes how much the density varies along a trajectory relative to the large-scale mean density $\langle \rho \rangle$. The right panel shows a related quantity which we call the \textit{local density variation}, $\sigma_\rho$, defined in Eq.~\eqref{eq:sigmarho}, which is the typical size of density fluctuations relative to the mean along that particular trajectory. In both cases, the solid white line corresponds to the median over $10^5$ random trajectories along 6 mpc, while the dashed lines enclose the 16th to 84th percentiles of the distribution. The distribution itself is normalised at each grid point and indicated by the purple shading.}
    \label{fig:var}
\end{figure*}

Apart from the statistics of minivoids and their profiles, the energy distribution within the simulation volume also provides further  insights. 
We are interested to estimate the degree of variance in the typically-observed dark matter density, in order to guide experiments that will be measuring over long timescales. 
However, our smallest discretisation scale $\Delta_x=0.4$ mpc (for $L=0.2$ pc, $N=512$) is still larger than the distance $d_{\rm year}\simeq 0.2$~mpc travelled by the Solar System around the Milky Way halo in one year at a speed of $v\sim 220$ km/s. 
This implies that, with our current simulation resolution, we cannot resolve the density variations for observation times under one to two years. 
Nevertheless, we have analysed $\mathcal{O}(10^5)$ randomly selected trajectories in the final snapshot of the $L=0.2$~pc simulation over a distance of $11\Delta_x\simeq 4.5$ mpc. For each trajectory, we estimate the density variation as a function of the displacement $\Delta r$ from the starting point.
Figure~\ref{fig:var} shows two measures of the density variation relevant  for different types of experiments relying on measurements over long periods of time and on individual short-term measurements. 

The first measure of the density variation, shown in the left panel of Fig.~\ref{fig:var}, is what we call the \textit{global density variation}, and measures how much the density contrast varies along a trajectory, relative to the average dark matter density over the whole simulation volume $\langle \rho \rangle$, 
\begin{equation}
    \Delta\delta(r)= \frac{\max{\rho(r)} - \min{\rho(r)}}{\langle \rho \rangle}\,.
    \label{eq:deltadelta}
\end{equation} 
As can be seen in Fig.~\ref{fig:var}, while the typical density sampled by an experiment at any one time is about 10\% of the average, the number can in fact be expected to vary between $\sim$5\% and $\sim$15\% over an ${\cal O}(1)$~year timescale. For longer times, i.e., $>10$ years, the variation can grow even more. We will refer to this result again in the next section. A varying density could potentially challenge the interpretation of multiple experiments attempting to observe and then test a putative axion signal over several short measurements at different points in time.

The second measure, shown in the right panel of Fig.~\ref{fig:var}, is a similar quantity which we simply call the \textit{local density variation}, and quantifies how the typical size of the density fluctuations changes relative to the average density {\it along a particular trajectory},
\begin{equation}\label{eq:sigmarho}
    \sigma_\rho(r) = \frac{1}{\langle \rho(r) \rangle_{\rm traj}} \sqrt{\frac{1}{n_r}\sum_{i = 1}^{n_r}(\rho(r_i)-\langle \rho(r) \rangle_{\rm traj})^2} \,.
\end{equation}
Here, $n_r$ denotes the number of grid points making up the trajectory $r$, and we use $\langle \rho (r) \rangle_{\rm traj}$ to make clear that this is the mean density along that trajectory.  
We define this measure as a standard deviation---rather than as a range as in Eq.~\eqref{eq:deltadelta}---because we use this in the context of single long measurements, though this choice is arbitrary.  
By this measure, the right panel of Fig.~\ref{fig:var} shows that the measured density can be expected to vary by 20--30\% initially for an observing time of $\mathcal{O}(1)$ year, and by almost 100\% if the measurement is carried out over even longer timescales.
This measure is consistent with the previous one, but the interpretation is subtly different, in that we must imagine here a continuous measurement of the axion field lasting long periods of time. This variation would affect the reconstruction of the axion's properties by an amount that goes roughly as the square root of the typical scale shown by e.g.~the white line in the right panel of Fig.~\ref{fig:var}. We discuss the implications of this result further in Sec.~\ref{sec:haloscopes}.

\subsection{Minivoid survival}

With current numerical methods we are not able to extend the simulations after $z\sim 100$. Indeed, it is unlikely that any N-body-based simulation that actually resolves individual miniclusters could ever be evolved to the present day on their own, due to the extreme computational demand placed on the resolution and box size. However, we can attempt to qualitatively describe what happens at later stages in the evolution.

To simplify the discussion, minivoids should be thought of as simply the space not occupied by miniclusters rather than actual entities in and of themselves, so what happens to them is necessarily dictated by whatever happens to the miniclusters. The only processes we can think of that can efficiently disrupt or affect the miniclusters are the tidal interactions with dense objects like stars. If, say, a minicluster had enough energy injected in a close encounter for the axions to become totally unbound, then the axion minicluster would grow into a long stream, spilling its mass into a volume given roughly by $\sim R^2 \sigma t$, where $R$ is the radius of the minicluster, $\sigma$ is its velocity dispersion, and $t$ the time since disruption. This is probably best thought of as the formation of a distinct type of object, a \emph{stream}~(see, e.g., ~\cite{OHare:2017yze,Foster:2017hbq,Knirck:2018knd}), but effectively what has happened is that the minivoids have partially been re-filled by axions. On the other hand, one of the key observations from our study is that the miniclusters and minivoids have stopped growing by the end of our simulations, with the void the densities approaching a constant value (see e.g. Fig. \ref{fig:void} on the right). Our expectation, as we will argue, is therefore that the stellar and tidal interactions, would likely only \textit{increase} the fraction of axions outside of miniclusters. An understanding of how the phase space distribution becomes re-filled, and the role of the minivoids and their substructure at $z=0$ is beyond our current scope and is left for future works.

\section{Implications for haloscopes}\label{sec:haloscopes}

\begin{figure}
    \centering
    \includegraphics[width=0.48\textwidth]{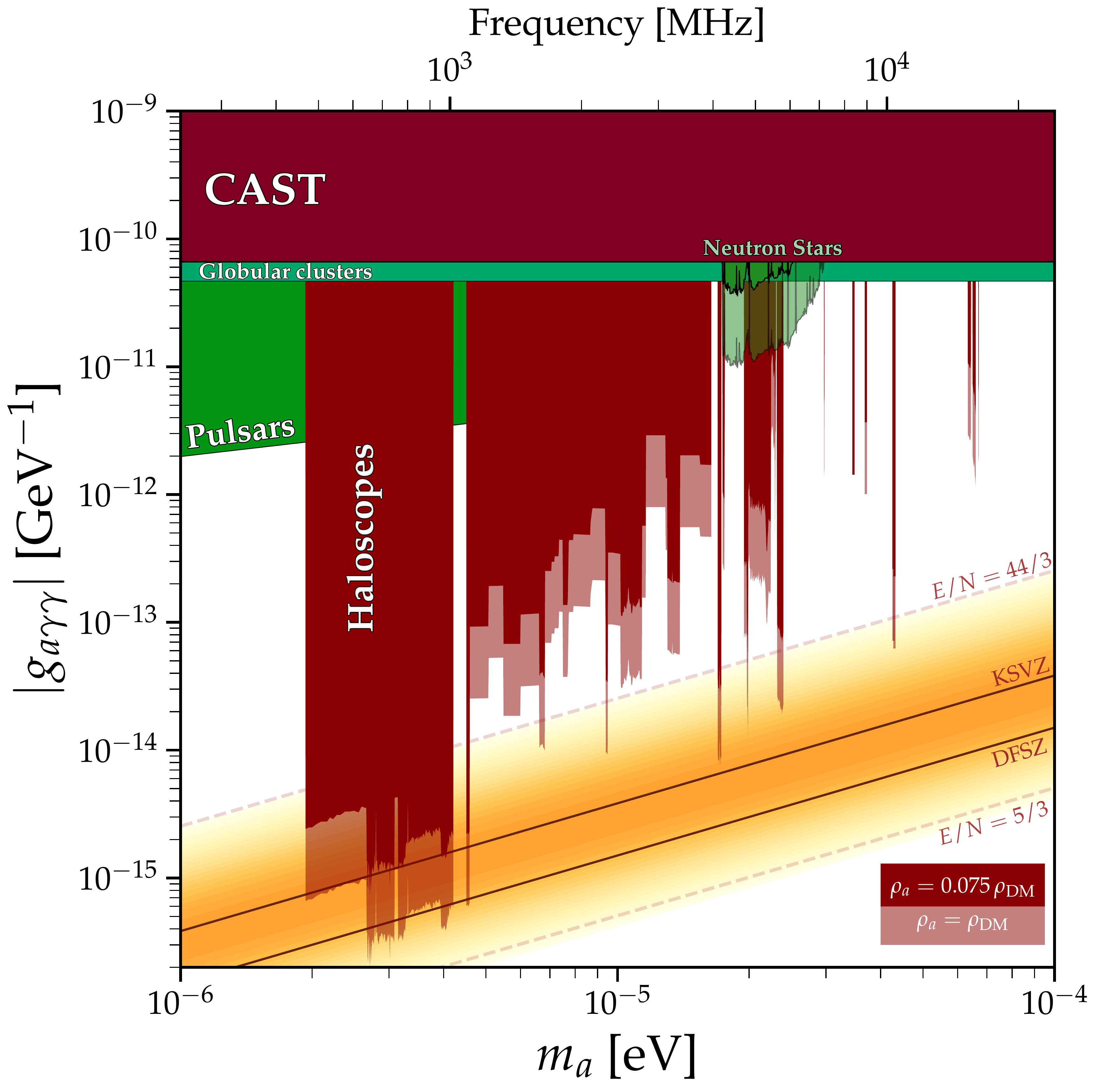}
    \caption{Current constraints on the axion-photon coupling in the $\upmu$eV--meV mass window, for two different assumptions about the ambient density of axions. In the lighter shades, we show the standard assumption that axions make up 100\% of the measured galactic dark matter distribution in the solar neighbourhood of $\rho_{\rm DM} = 0.45$~GeV~cm$^{-3}$. The darker shade bounds our estimate of the lowest assumption for the typical ambient density of dark matter in the worst-case scenario that the local density of axions is primarily inside of miniclusters, yet experiments probe the space between miniclusters. The red haloscope bounds are taken from Refs.~\cite{Asztalos2010,ADMX:2018gho,ADMX:2019uok,ADMX:2021nhd,ADMX:2018ogs,Bartram:2021ysp,Lee:2020cfj,Jeong:2020cwz,CAPP:2020utb,Lee:2022mnc,Yi:2022fmn,Adair:2022rtw,Grenet:2021vbb,HAYSTAC:2018rwy,HAYSTAC:2020kwv,McAllister:2017lkb,Quiskamp:2022pks,Alesini:2019ajt,Alesini:2020vny,CAST:2020rlf,DePanfilis,Hagmann,TASEH:2022vvu,TASEH:2022noe}, CAST from Ref.~\cite{CAST:2007jps,CAST:2017uph}, globular cluster stellar cooling bound from Ref.~\cite{Dolan:2022kul}, a search for axions produced in pulsar polar cap cascades~\cite{Noordhuis:2022ljw}, and the neutron star dark matter bound from Ref.~\cite{Foster:2022fxn}. Since the last constraint also relies on the assumption that the axions make up galactic dark matter, we have also applied the rescaling factor as with the haloscopes. Limit data and plots available at Ref.~\cite{AxionLimits}.}
    \label{fig:axionphoton}
\end{figure}

Having quantified the extent to which the typical local density of axions in a galactic halo is suppressed due to the presence of minivoids, we now wish to evaluate the implications for efforts to search for the dark-matter axion in laboratory experiments. For illustrative purposes, we shall focus on experiments utilising the axion's coupling to the photon, $g_{a\gamma}$, as these are the subject of the most experimental activity. In principle, though, our results apply to all direct searches for axions via any other coupling, as long as the post-inflationary scenario is true and the resulting miniclusters are present in our universe---this typically will be the case for axions in the mass range $\sim 10\upmu$eV--meV.%
\footnote{Axion models with domain wall number greater than 1 may have qualitatively different distributions of miniclusters and minivoids, which is deserving of a dedicated study} 
Hence, experiments like CASPEr-electric~\cite{JacksonKimball:2017elr}, as well as the proposals of Refs.~\cite{Berlin:2022mia, Arvanitaki:2021wjk}, which purport QCD axion sensitivity in the same mass range via the electric-dipole-moment coupling may also be impacted. 

Essentially, all direct searches for axions are based on a model for the local behaviour of the field in terms of coherent oscillations,
\begin{equation}\label{eq:axionfield}
    a(t) \approx \frac{\sqrt{2\rho_a}}{m_a} \cos{(\omega t + \phi)},
\end{equation}
where $\phi$ is an arbitrary phase. We ignore all spatial dependence here as it is unimportant for this discussion. The frequency of the oscillations is given by the kinetic energy in the field $\omega \approx m_a(1+v^2/2)$, where $v$ is the dark matter speed in units of $c$. Since the dark matter speed is only a correction of $v \sim 10^{-3}$ and varies with a dispersion, $\sigma_v$, of approximately the same scale, this implies the field oscillates coherently over $m_a/m_a \sigma_v^2 \sim 10^6$ cycles. The velocity will be drawn from the local velocity distribution, however, so when the field is observed for timescales longer than $10^6/m_a$, any resulting signal that is tied to those oscillations will be distributed in frequency according to the same distribution up to a change of variables---this, in haloscope jargon, is called the lineshape.

The majority of axion experiments (and all axion experiments using the photon coupling) couple linearly to the axion field, e.g., $\mathcal{L} \sim g_{a\gamma} a F \tilde{F}$. This means that signals will scale as $a ^2 g^2_{a\gamma}$. Ignoring $\mathcal{O}(1)$ factors specific to individual experiments, the generic scaling of electromagnetic signal power on the quantities of interest is
\begin{equation}
    P(\omega_i) \propto g^2_{a\gamma} \alpha(\omega_i) \rho_a f(\omega_i) \, ,
\end{equation}
where $f(\omega_i)$ is the axion lineshape binned at some arbitrary set of frequency bins $\omega_i$. The amplitude $\alpha(\omega_i)$ is a stochastic quantity drawn from a Rayleigh distribution~\cite{Foster:2017hbq} that is present if the phases in Eq.~\eqref{eq:axionfield} are randomly drawn in every coherence time (this may not be the case in the vicinity of miniclusters; however, the conservative approach is to assume that it is).

From here we see that experimental sensitivity to $g_{a\gamma}$ scales as $1/\sqrt{\rho_a}$, implying a factor of $\sim$3 suppression in limits when comparing typical minivoid densities to the baseline homogeneous assumption
that the observable axion density is exactly equal to the astronomically inferred dark matter density in the solar neighbourhood. Estimates of the latter vary in the literature---in fact, there is a long history of these types of measurements~\cite{Read:2014qva}---but most recent analyses using \textit{Gaia} data place the local density of dark matter in the range $\rho_{\rm DM} = [0.2,0.7]$~GeV~cm$^{-3}$~\cite{deSalas:2020hbh}. The most important type of measurement for our purposes are the local techniques which aim to solve the Jeans equation for tracer populations of stars in the solar neighbourhood. Then after subtracting the baryonic contribution to the gravitational potential, one can infer the (sub-dominant) contribution coming from dark matter. These also usually give values in the same range, but importantly rely upon stellar tracer populations that span at least a few $\sim$100~pc. This is far, far above the scale probed by experiments, which is less than a milliparsec.

The results of the previous section suggest that the typical density sampled by an experiment at a given instant is around 10\% of the average density in the box which we use as a proxy for the distribution of axions in a similar-sized volume within our galaxy. The convention in the axion community has been to adopt the value $\rho_{\rm DM} = 0.45$~GeV~cm$^{-3}$ and to assume 100\% of dark matter is in the form of detectable axions in the ambient dark matter in the solar neighbourhood. In Fig.~\ref{fig:axionphoton} we show the status of experimental limits on $g_{a\gamma}$ in the approximate mass window relevant to the post-inflationary scenario. The standard published limits under the dark matter density convention mentioned above are shown in lighter-shaded colours. We show the extent to which this sensitivity is reduced in the \textit{worst-case scenario} in which the volume in our simulation represents the typical density inside a galaxy at $z=0$. A $\sim$10\% reduction in the typical density corresponds to the factor of $\sim$3 suppression in sensitivity. Interestingly---or perhaps worryingly---this is comparable to the ratios in the couplings of the two common axion model benchmarks, the KSVZ~\cite{Kim:1979if, Shifman:1979if} and DFSZ~\cite{Dine:1981rt, Zhitnitsky:1980tq}  models: KSVZ/DFSZ $\sim 1.92/0.75 \sim 2.56$. Hence, we can make the statement that an experiment ruling out the DFSZ model over some mass range will still have ruled out KSVZ even when most of the axions are bound up in miniclusters. 

We generally expect from the plateauing growth of the miniclusters at the latest redshifts that the voids should not lose much more density in axions. Because of this fact, as well as the fact that ambient density ought to be partially refilled by disrupted miniclusters, we may expect our estimate of the typical density to be a conservative lower bound. It could prove to be higher once the galaxy has formed in full, and as the miniclusters are disrupted with their contents virialised into the main host. However, previous studies have shown that something around 60-70\% of miniclusters orbiting around the solar neighbourhood would remain intact~\cite{Kavanagh:2020gcy, Shen:2022ltx}.

This suppression in density is perhaps most impactful for resonance-based experiments like ADMX~\cite{ADMX:2021nhd}, CAPP~\cite{Lee:2022mnc}, HAYSTAC~\cite{HAYSTAC:2020kwv}, TASEH~\cite{TASEH:2022vvu, TASEH:2022noe}, GrAHal~\cite{Grenet:2021vbb}, RADES~\cite{CAST:2020rlf}, ORGAN~\cite{McAllister:2017lkb, Quiskamp:2022pks} and MADMAX~\cite{TheMADMAXWorkingGroup:2016hpc}, that rely on only short total integration times at any one frequency. The suppressed typical density reduces the available signal these experiments can observe, and hence should be taken seriously when considering how much of the axion parameter space in the post-inflationary mass range they have really constrained.

In contrast, for experiments that do not rely on resonances, and instead operate in a broadband mode over a wide range of frequencies, our aim was to study the possible implications for (i)~individual measurements at different times spanning a relatively long period, and (ii)~continuous measurements for a long period of time. 
In the first context (left panel of Fig.~\ref{fig:var}.), the most relevant implication of this result is for multiple different experiments each testing the veracity of a putative signal at different points in time. Na\"{\i}vely, we would expect that if a signal were real it would remain at the same strength at every given measurement. Here, we see that measurements can be expected to vary in strength over several-year periods. Furthermore, to reconstruct the value of the axion's coupling, one has to break the degeneracy $g_{a\gamma}^2 \rho_a$ somehow, usually by assuming a value of $\rho_a$. We have shown in this work that a 10\% suppression compared to the measured local dark matter density is expected, although this suppression could vary between, say, 5\% and 30\% on a timescale of only a couple of years.

In the second context---i.e., the right panel of Fig.~\ref{fig:var}---we computed not the range in density suppressions, but rather the standard deviation in densities along many different trajectories. As in this case, we take the density relative to the mean density at that specific location, it is more relevant for experiments conducting single continuous measurements. This is not a relevant question for many current haloscopes, however for upcoming broadband haloscopes, it is. The best example is BREAD~\cite{BREAD:2021tpx} which does not attempt any resonant enhancement of the axion signal, but rather has a broadband acceptance of photons over a wide band of frequencies. To compensate for its lower sensitivity at any given mass, it instead looks for signals over much longer integration times. BREAD is additionally relevant as in its proposed guise of ``GigaBREAD'' in which a coaxial horn antenna will be used as the detector, it will have sensitivity to axions squarely in the post-inflationary mass range. Projections published by the collaboration assume total data-taking times ranging from 10 to 1000 days. Taking the longer of those durations, we can inspect the right panel of Fig.~\ref{fig:var} to see that a signal variation of $\sim 30\%$ is \textit{expected} within that time. 

\section{Conclusions}\label{sec:conc}

In this article, we have presented the results from a set of lattice and N-body simulations for the post-inflationary QCD axion. We performed for the first time a realistic simulation of the axion field from the epoch of cosmic strings before the collapse $z\sim 10^{16}$ and the time when the field configuration collapses into halos and forms minivoids, $z\sim \mathcal{O}(10^2)$. This was done by combining a three-stage simulation, with lattice methods to solve the axion relativistic field equation and the nonrelativistic Schr\"odinger-Poisson system, and N-body methods for the collisionless dark matter dynamics. 

We have placed a particular emphasis on the axions that occupy minivoid regions, in the space between small miniclusters and larger minicluster halos. As the minivoids occupy the vast majority of the sub-pc-sized simulation volume, the same will be true of a typical place in our galaxy. Direct detection experiments on Earth therefore do not sample the value of the dark matter density of $\sim 0.4$~GeV~cm$^{-3}$---which is obtained through measurements of stellar dynamics on scales well above this---but rather a suppressed value, given by the typical axion density in the voids. The quantitative answer to how great this suppression is, is around 10\% of that density. We have tested various differences in the technical configuration of the simulation, including the physics included in the evolution, as well as the initial conditions, and find that this number is relatively robust. 

We also made an estimate of the variance in this density, arguing that the density of axions along typical $\mathcal{O}$(1)-year-long trajectories through the Galaxy can be expected to vary quite considerably. While the 10\% suppression relative to the global average on galactic scales does not vary much, for experiments like broadband haloscopes making single continuous measurements of a potential axion signal, the density should be expected to vary by several tens of percent.

We should finish by remarking that our results strictly only apply at the final redshift of our simulations---typically $z_f\gtrsim 100$---which is of course well before the present day. Extrapolating our result to $z=0$ is beyond the scope of this work; however, we have tried to make statements that are conservative in several ways. 

Firstly, it should be said that the fractions of axions in miniclusters and in voids are both plateauing by the end of the simulation, implying that the evolution is approaching a somewhat steady state. Secondly, the next important process for the lives of the miniclusters will be tidal disruption.  The case can be made that our estimation of the density suppression and its variation can both be taken as a conservative lower bound. Due to the tidal and stellar disruption of the miniclusters, the axion dark matter distribution could be described by numerous virialised streams of axions, leading to larger typical densities, but also larger variations. Disruption has been studied for applications for indirect signals, e.g., collisions between miniclusters and neutron stars~\cite{Edwards:2020afl}. However further study is needed to quantify how much our results might change after considering disruption in the case of direct detection. It may well be that sufficient disruption occurs for the haloscopes to reclaim much of their earlier assumed sensitivity.

\acknowledgments
BE acknowledges support from the Deutsche Forschungsgemeinschaft.
CAJO is supported by the Australian Research Council under grant number DE220100225.
GP is supported by a UNSW University International Postgraduate Award. 
JR is supported by the grant PGC2018-095328-B-I00(FEDER/Agencia Estatal de Investigacion) and FSE-DGA2017-2019-E12/7R (Gobierno de Aragon/FEDER) (MINECO/FEDER), the EU through the ITN Elusives
H2020-MSCA-ITN-2015/674896 and the Deutsche Forschungsgemeinschaft under grant SFB-1258 as a Mercator Fellow. 
Y$^3$W is supported in part by the Australian Government through the Australian Research Council’s Future Fellowship (project FT180100031).
This research was undertaken using the computational cluster \emph{Katana}~\cite{katana} supported by Research Technology Services at UNSW Sydney, the \emph{Gadi} cluster from the National Computational Infrastructure (NCI) and supported by the Australian Government, and the HPC system \emph{Raven} at the Max Planck Computing \& data facility (MPCDF). We acknowledge the use of the \texttt{\href{https://github.com/franciscovillaescusa/Pylians3}{Pylians}}~\cite{Pylians} library for the analysis and \codo{matplotlib}~\cite{Hunter:2007}, \codo{cmasher}~\cite{cmasher} and \codo{\href{https://yt-project.org/}{yt}}~\cite{yt} for visualisation.

\appendix

\section{Details of numerical methods}\label{app:methods}

In this appendix, we discuss the technical details of our lattice and N-body simulations.

\subsection{Lattice simulations}

As in Refs.~\cite{Vaquero:2018tib,OHare:2021zrq} we use the public code \texttt{jaxions}\footnote{\href{https://github.com/veintemillas/jaxions}{https://github.com/veintemillas/jaxions}.} to simulate axions around the QCD phase transition. In particular, we consider the Klein-Gordon evolution equation for the complex field $\phi$, 
\begin{equation}
    \ddot{\phi}+3H\dot{\phi}-\frac{\nabla^2}{R^2}\phi+\lambda\phi(\vert\phi\vert^2-f_a^2)=0,\label{eq:kgfull}
\end{equation} 
which has as its solutions radial and massive \emph{saxion} modes, angular and massless \emph{axion} modes, and strings. 
Equation~(\ref{eq:kgfull}) is used to evolve $\phi$ from an initial redshift of $z_i\sim 10^{16}$ until the collapse of topological defects that happens around $z\sim 10^{13}$. From there on, only the angular degree of freedom needs to be considered, and  
the equation of motion reduces to
\begin{equation}
    \ddot{a}+3H\dot{a}-\frac{\nabla^2}{R^2}a+m^2_a(T)f_a\sin\left(\frac{a}{f_a}\right)=0\,,
    \label{eq:kgax}
\end{equation} 
where $a/f_a=\arg\phi$, and $m_a(T)\propto T^{-n}$ is the temperature-dependent axion mass with power-law index $n=7$~\cite{Borsanyi:2016ksw,Vaquero:2018tib}. 

We solve Eqs.~\eqref{eq:kgfull} and \eqref{eq:kgax} using a finite difference method and the symplectic integrator RKN4~\cite{1992}. 
The spatial derivatives are calculated from two neighbouring points and accurate to $\mathcal{O}(\Delta^4_x)$, where $\Delta_x=L/N$, while the time-step is dynamically adjusted to resolve the fastest modes in the box~\cite{Vaquero:2018tib}. 
To solve Eq.~\eqref{eq:kgfull} we implement the so-called \emph{fat-string} or Press-Ryden-Spergel (PRS) trick~\cite{PRS,Moore:2001px} that keeps the string width constant in comoving coordinates by rescaling the potential coupling $\lambda\to\lambda/R^2$. A resolution of $N^3=8192^3$ grid points is required for simulations in boxes of side length $L>8L_1$, in order to avoid the unphysical destruction of the string-wall network. 
For smaller simulation volumes a resolution of $N^3=4096^3$ is adopted. After a few background field oscillations, the axion field $a(x)$ enters the linear regime as the field values become smaller. An adiabatic approximation is applied at $z\sim 5\times 10^{11}$, well after the axitons are expected to disappear as the axion mass reaches its zero-temperature value. 

\begin{figure}
    \centering
    \includegraphics[width=0.9\columnwidth]{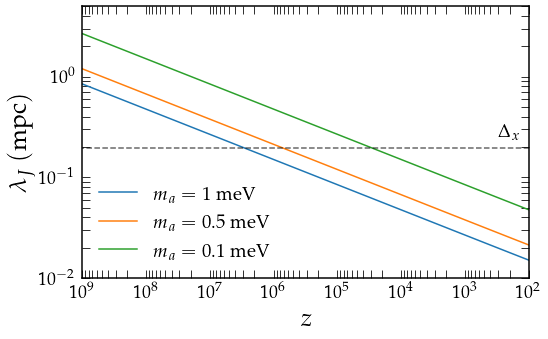}
    \caption{Axion comoving Jeans wavelength $\lambda_J$ as defined in Eq.~\eqref{eq:laJ} as a function of redshift for different axion masses $m_a$.  The grid resolution in our Schr\"odinger-Poisson simulations, $\Delta_x$, is indicated by the horizontal dashed line. }
    \label{fig:lambdaJ}
\end{figure}

At this point, we apply the nonrelativistic approximation and map the conformal axion field $a\tau$, where $\tau$ denotes the conformal time $\mathrm{d}\tau = \mathrm{d}t/a$, into the complex conformal field $\psi$ as in Eq.~\eqref{eq:paxion}. This is done by identifying the real and imaginary parts of $\psi$, i.e.,
\begin{align}
    {\rm Re}[\psi]& = \frac{\sqrt{2\tau m_a}}{2}a\tau , \\ 
    {\rm Im}[\psi]& = \frac{\sqrt{2\tau m_a}(\n+2)R'}{8\tau m_a R}a\tau +\frac{1}{\sqrt{2\tau m_a}}(a\tau)'.
\end{align} 
Note that, although this identification effectively means that we double the degrees of freedom stored in the field, the system of equations, Eqs.~\eqref{eq:sp1} and \eqref{eq:sp2}, that describes the evolution of $\psi$ is a first-order system.  Thus, the computational cost required to solve them is the same as that needed to solve the second-order Eq.~\eqref{eq:kgax}. We solve the Schr\"odinger-Poisson system with a resolution of $1024^3$ grid points and similar schemes as described above, i.e., a sympletic time integration and finite difference for the spatial derivatives, also implemented in \codo{jaxions}. The effects of gravity in the Schr\"odinger-Poisson system enter via the Madelung transformation in the equation of motion for the phase $S = \mathrm{arg}\psi$, 
\begin{equation}
    S'=\frac{1}{2m_a}\left(\frac{\nabla^2\vert\psi\vert}{\vert\psi\vert}-(\nabla S)^2\right)-m_a\Phi_N,
\end{equation} 
while the equation of motion for the modulus $\vert\psi\vert$ can be recast into the energy density continuity equation. The phase field $S$ will then drive the axion density towards the minima of the potential $\Phi_N$, and itself starts to wrap the fundamental domain $[0,2\pi)$ according to the depth of the well. Thus, the numerical evaluation will lose its accuracy when $2\pi$ phase jumps occur near the discretisation scale. In other words, there is a maximum value that can be reached in the gradient of $S$ (i.e., the velocity field), 
\begin{equation}
    (\nabla S)_{\rm max}\sim \frac{\pi}{\Delta_x}\,,\label{eq:cond}
\end{equation} 
which limits the evolution of the Schr\"odinger-Poisson system.
This maximum phase gradient can be related to the maximum difference in the values of the gravitational potential that can be read at each time step. The condition~\eqref{eq:cond} is reached at redshift $z_{*}\sim 2\times 10^6$, after which we continue the evolution of the system using N-body simulations,  since with our choice of $m_a$ the Jeans wavelength is not resolved anymore, $\lambda_{\rm J}<\Delta_x$; see Fig. \ref{fig:lambdaJ}.

\subsubsection{A note on velocities}
As long as the particles are in the single-streaming regime, we can assign a single velocity at each local point by solving the Euler equation 
\begin{equation}
    \nabla\cdot \mathbf{v}(x) = \frac{\nabla^2 S(x)}{m_{a}}\,,
\end{equation} 
and hence
\begin{equation}
    \mathbf v(x) = \frac{\nabla \arg(\psi)}{m_a}.
\end{equation} 
This task can be performed using the full complex field~$\psi$ by exploiting the relation 
\begin{equation}
    \nabla S = {\rm Im}\left(\frac{\psi^*\nabla\psi}{\vert\psi\vert^2}\right)=\frac{\psi_r\nabla\psi_i-\psi_i\nabla\psi_r}{\psi_r^2+\psi_i^2},
\end{equation} 
where $\psi_{r,i}$ denote the real and imaginary parts, respectively. Typical velocities in the initial conditions are the order $\langle v\rangle \sim 10^{-9}$ in natural units.

\subsection{N-body simulations}

Given the large range of time evolution, an N-body run is costly in terms of computational resources and runtime. We therefore limit the resolution of our simulations in this work to $512^3$ dark matter particles and leave higher resolution runs for a future publication. 
For the simulations with box sizes $L=8L_1$ and $L=16L_1$, we use a TreePM method as implemented in \texttt{GADGET-4}, with a PM grid of $512^3$ points and a numerical softening length of $\ell_s\simeq 1.95 \times 10^{-5}~{\rm pc}/h\simeq 4~{\rm AU}/h$ in comoving units. This value is $\sim 30$ and $\sim 60$ times smaller than the average particle separation in the two simulation volumes used in this study.
We modified \texttt{GADGET-4} to include the effects of radiation that dominates the energy budget at the initial time and perform our simulations using the following cosmological parameters: $\Omega_m=0.3, \Omega_{\Lambda}=0.7, \Omega_r = 8.49\times 10^{-5}$, and $h=0.71$. 
The average particle mass in our simulations is $m_{\rm av}=1.15\times 10^{-16} M_{\odot}/h$ and $m_{\rm av}=4.4\times 10^{-17} M_{\odot}/h$, respectively, in our $L=0.2$ pc and $L=0.4$ pc simulations.

\section{More on minivoids}\label{app:void}

\begin{figure}
    \centering
    \includegraphics[width=0.45\textwidth]{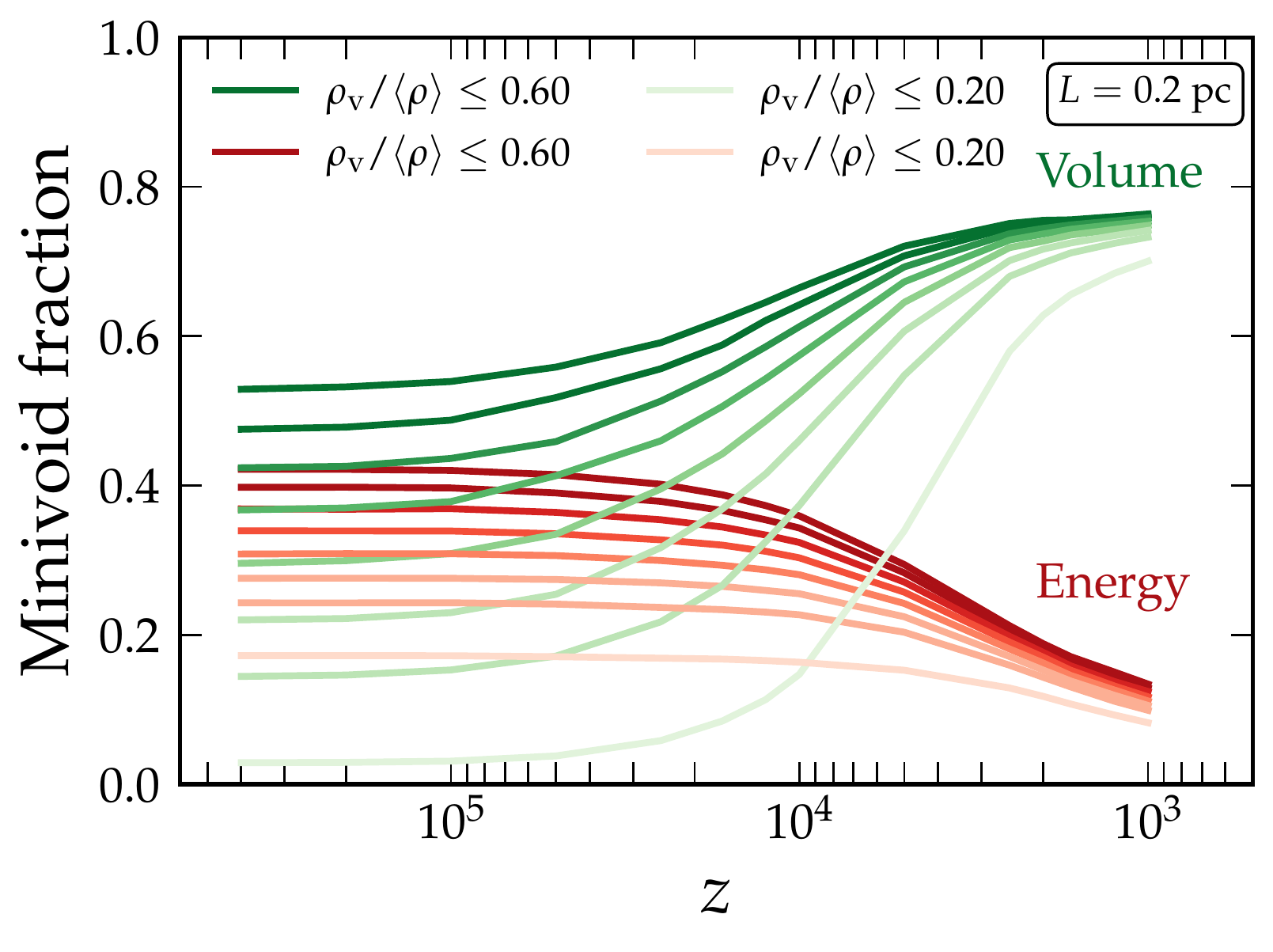}
    \caption{Minivoid volume fractions $f_{\rm v}$ (green) and typical energy densities $\rho_{\rm v}/\langle\rho\rangle$ (red) as a function of redshift. We highlight how results at late times are independent of the threshold parameter of the void finder; see Sec.~\ref{sec:finder} for details.}
    \label{fig:void8}
\end{figure}

\begin{figure}
    \centering
     \includegraphics[width=0.45\textwidth]{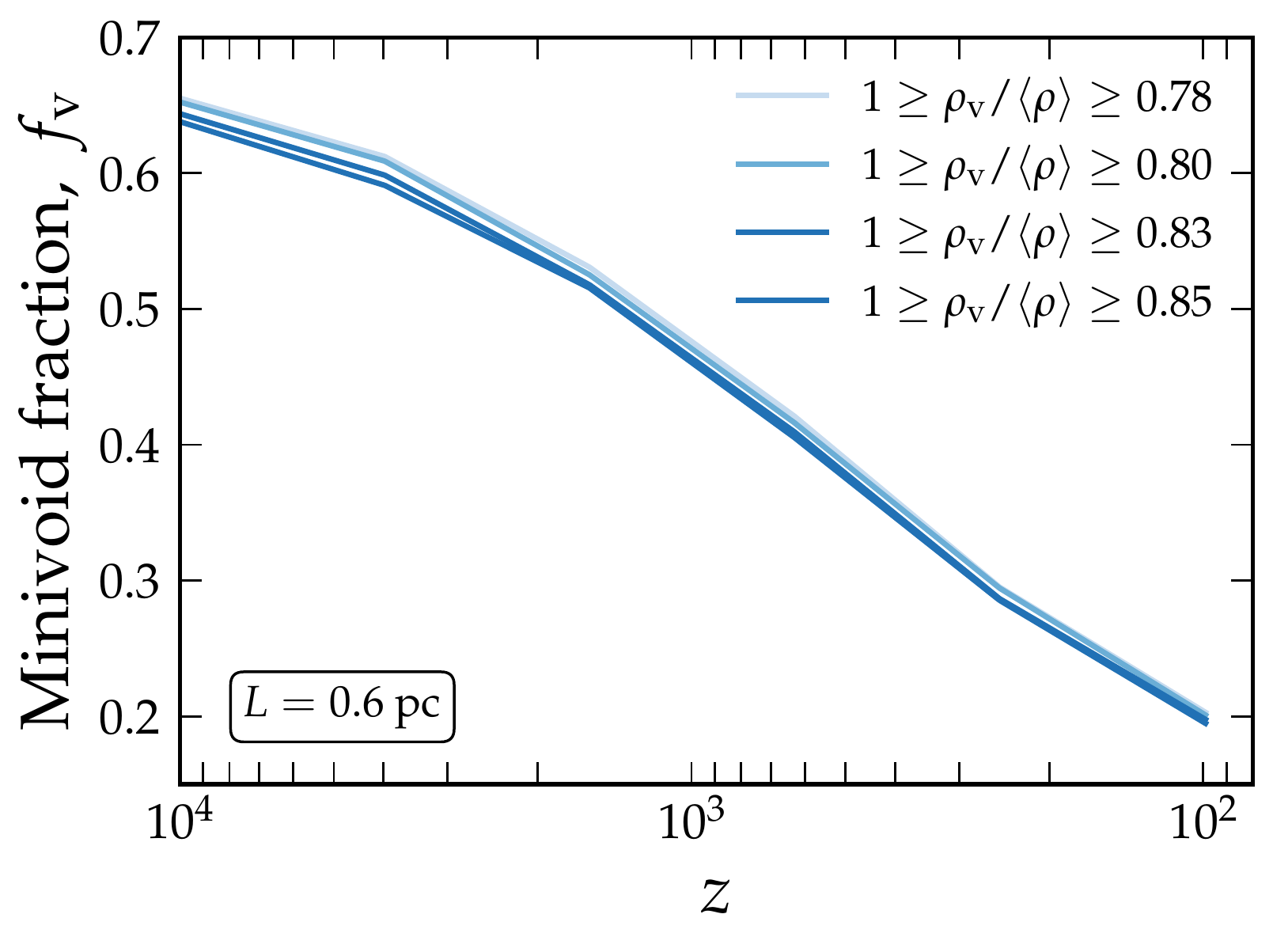}
    \caption{Minivoid volume fractions $f_{\rm v}$ for ``mild'' underdensities, defined as regions with $\sim 80\%$ of the average value $\langle\rho\rangle$ over the simulation volume. At the final simulation time of the $L=0.6$~pc run (i.e., $z_f=99$), these mildly underdense regions occupy $\sim 20\%$ of the simulation volume.}
    \label{fig:fila}
\end{figure}

As described in the main text,  minivoids are identified based on a pre-determined density threshold $\delta_a^{\rm thr}$ on the axion energy density built from the particle positions. We show in this appendix that the resulting void statistics do not depend on the specific choice of $\delta_a^{\rm thr}$. 

Figure~\ref{fig:void8} shows the minivoid volume fraction $f_{\rm v}$ and energy density $\rho_{\rm v}/\langle\rho\rangle$ extracted from the $L=0.2\,\mathrm{pc}$ simulation using a range of threshold values $-\delta_a^{\rm thr}=\{0.4,0.45,0.5,0.55,0.6,0.65,0.7,0.8\}$.
Evidently, both $f_{\rm v}$ and $\rho_{\rm v}/\langle\rho\rangle$ converge to the values given in Eq.~\eqref{eq:finalvoid} for all thresholds $\delta_a^{\rm thr}\geq-0.7$. We find deviations for the choice of $\delta_a^{\rm thr}=-0.8$, as minivoids with $\sim 20\%$ energy of the overall average value $\langle\rho\rangle$ occupy a sizeable fraction of the box. 
This behaviour is also observed in the $L=0.4,0.6$~pc simulations.

\begin{figure}
    \centering
    \includegraphics[width=0.45\textwidth]{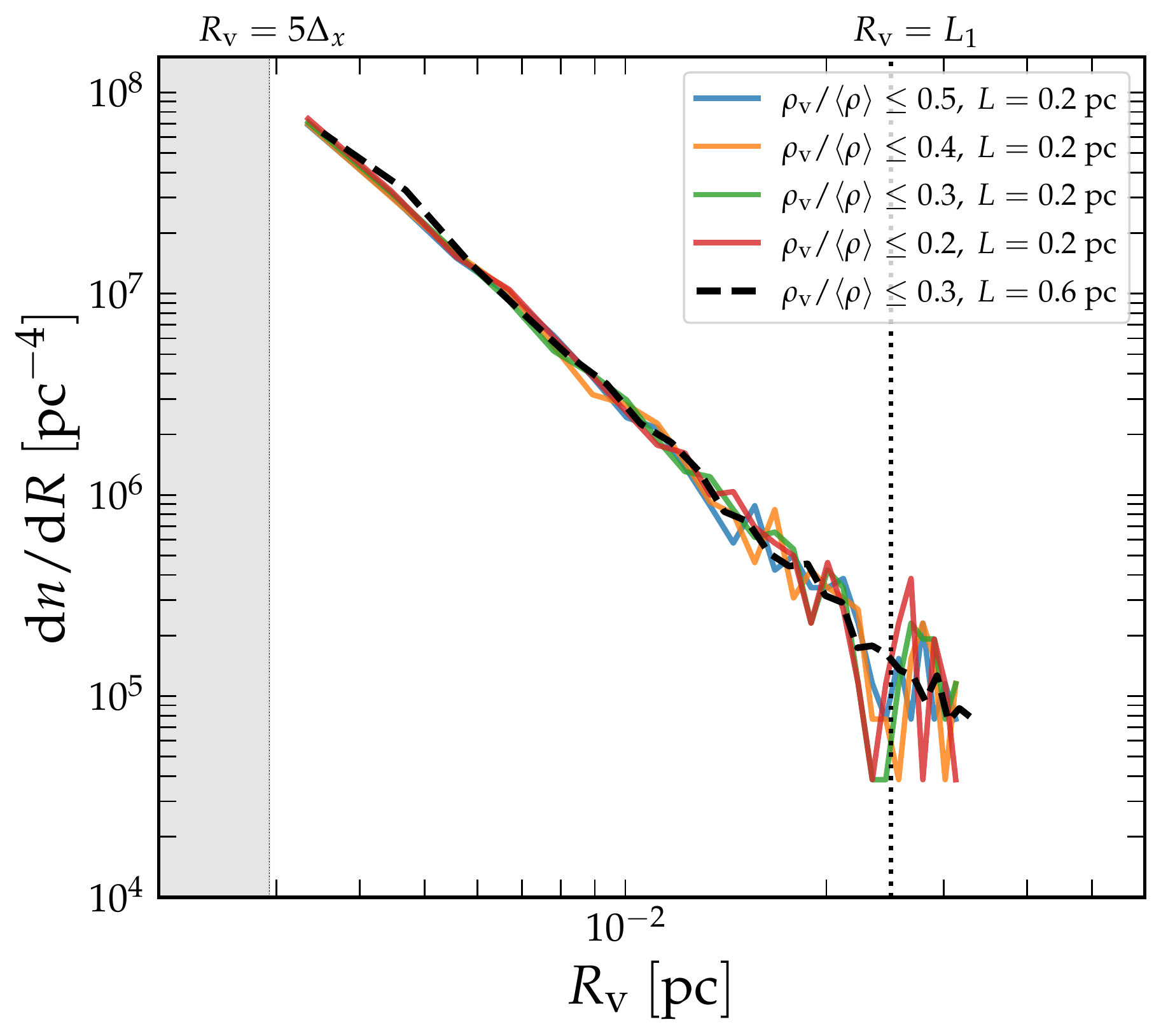}
    \caption{Minivoid size function extracted from the $L=0.2$~pc simulation at $z=999$ for various choices of the void threshold. 
    For reference, the black dashed line corresponds to the $L=0.6$~pc simulation result at the same redshift.}
    \label{fig:vsf_size}
\end{figure}

In addition, we want to stress that, while $80\%$ of the volume is occupied by low-density minivoids, the remaining $\sim 20\%$ of the volume is occupied by regions that are only mildly underdense---usually in the form of filaments connecting minicluster halos---and by regions of average density that limit the boundaries of miniclusters. We have checked this by imposing a \emph{lower} density threshold in the void finder algorithm, i.e.,
\begin{equation}
    1\geq \rho_{\rm v}/\langle\rho\rangle \geq \delta_a^{\rm thr}. 
\end{equation} 
Figure~\ref{fig:fila} shows the resulting $f_{\rm v}$ from the $L=0.6$~pc simulation  for several choices of $\delta_a^{\rm thr}$.

Finally, Fig.~\ref{fig:vsf_size} shows the void size function at $z=999$ for various choices of minivoid density thresholds for two different simulation volumes. Evidently, they are in good agreement with each other. Note that the void size function does not change substantially between $z=999$ and $z=99$ (see Fig.~\ref{fig:vsf}).

\bibliography{axions.bib}
\bibliographystyle{bibi}

\end{document}